\documentclass[preprint,review,12pt]{elsarticle}
\usepackage{amssymb}
\usepackage{amsmath,bm}
\usepackage{colortbl,booktabs}
\usepackage{tabularx}
\usepackage{threeparttable}
\usepackage{multicol,multirow}
\usepackage{subfigure}
\usepackage [font=footnotesize,labelfont=bf]{caption}
\usepackage{rotating}
\usepackage{tikz}
\usepackage{hyperref}
\usepackage{lineno}
\journal{Elsevier}

\begin{document}
%%\linenumbers
\captionsetup[figure]{labelfont={bf},labelformat={default},labelsep=period,name={Fig.}}
\begin{frontmatter}

\title{Multiphase SPH for surface tension: resolving zero-surface-energy modes and achieving high Reynolds number simulations}

\author[address1,address2]{Shuaihao Zhang}
% \ead{szhang07@connect.hku.hk}
\author[address1]{Sérgio D.N. Lourenço}
% \ead{lourenco@hku.hk}
\author[address2]{Xiangyu Hu \corref{mycorrespondingauthor}}
\cortext[mycorrespondingauthor]{Corresponding author.}
\ead{xiangyu.hu@tum.de}

\address[address1]{Department of Civil Engineering, The University of Hong Kong, Pokfulam, Hong Kong SAR, China}
\address[address2]{School of Engineering and Design, Technical University of Munich, 85748 Garching, Germany}

\begin{abstract}
	This study introduces a Riemann-based Smoothed Particle Hydrodynamics (SPH) framework for the stable and accurate simulation of surface tension in multiphase flows, with density and viscosity ratios as high as 1000 and 100, respectively. 
	The methodology begins with the computation of surface stress, from which surface tension is derived, ensuring the conservation of momentum. 
	For the first time, this study identifies the root cause of particle disorder at fluid-fluid interfaces, attributed to a numerical instability defined herein as \textit{zero-surface-energy modes}. To address this, we propose a novel penalty force method, which eliminates zero-surface-energy modes and significantly enhances the overall stability of the simulation.
	Importantly, the penalty force correction term is designed to maintain momentum conservation.
	The stability and accuracy of the proposed framework are validated through several benchmark cases with analytical solutions, performed under both two-dimensional and three-dimensional conditions. 
	Furthermore, the robustness of the method is demonstrated in a three-dimensional high-velocity droplet impact scenario, achieving stable performance at high Reynolds numbers ($Re=10000$) and Weber numbers ($We=25000$). 
	To the best of our knowledge, this represents the first successful demonstration of a mesh-free method achieving stable multiphase flow simulations under such extreme $Re$ and $We$ conditions.
	A qualitative comparison with previous experimental results is also conducted, confirming the reliability of the simulation outcomes. 
	An open-source code is provided for further in-depth study.
\end{abstract}


\end{frontmatter}
%%%%%%%%%%%%%%%%%%%%%%%%%%%%%%%%%%%%%%%%%%%%%%%%%%%%%%%%%%%%%
%
% 1 Introduction
%
%%%%%%%%%%%%%%%%%%%%%%%%%%%%%%%%%%%%%%%%%%%%%%%%%%%%%%%%%%%%%
\section{Introduction}
\label{introduction}
Multiphase flow problems are widely encountered across a broad range of disciplines, including the interaction of water and oil in mechanical engineering, the transport and dispersion of pollutants in environmental engineering, and the mixing of water and gas in porous soil within geotechnical engineering. 
They are also crucial in biomedical fields, such as blood flow in capillaries and respiratory airflow involving liquid films. Furthermore, these flows play a significant role in industrial applications such as oil recovery, chemical manufacturing, food processing, and even in energy systems, such as multiphase flows in fuel cells and nuclear reactors.

The numerical simulation of multiphase flows is particularly challenging due to the sharp variations in density and viscosity that typically occur at the interfaces separating different fluids.
Surface tension effects play a crucial role in many multiphase flow phenomena, as they govern the dynamics at fluid-fluid interfaces and significantly influence processes such as droplet formation, coalescence, and breakup \cite{adami2010new, zhang2023improved,pozorski2024smoothed}.
Numerical methods for multiphase flow simulations can be broadly categorized into Eulerian methods and Lagrangian methods.
Eulerian methods, such as the volume of fluid (VOF) method \cite{scardovelli1999direct}, the level-set method \cite{sussman1994level}, the lattice Boltzmann method (LBM) \cite{chen1998lattice}, and front-tracking \cite{tryggvason2001front}, explicitly capture or track fluid interfaces using 
dedicated interface representing variables and algorithms. 
In contrast, fully Lagrangian particle-based methods naturally represent interfaces by following the motion of particles \cite{adami2010new, colagrossi2003numerical, yang2022consistent}. 
This adaptive representation enables the efficient handling of strong interface deformations, including breakup, by assigning particles to specific phases throughout the simulation.
As one of the most representative particle-based methods, the smoothed particle hydrodynamics (SPH) method  \cite{gingold1977smoothed, lucy1977numerical} has been extensively applied to multiphase fluid simulations, particularly in scenarios involving surface tension effects \cite{morris2000simulating, nugent2000liquid, tartakovsky2005modeling, hu2006multi, le2025smoothed}.

The continuum surface force (CSF) method, initially introduced by Brackbill et al. \cite{brackbill1992continuum}, is a widely adopted approach for modeling surface tension effects. 
This method converts surface tension into a continuous volumetric force distributed near the interface using a surface delta function.
Based on the surface curvature, normal vector, and delta function, this force is incorporated into the computational model as an external force term.
The CSF model was first introduced into the SPH framework by Morris \cite{morris2000simulating} and has since been widely applied \cite{adami2010new, zhang2023improved, blank2023modeling}.
However, the computation of interface curvature is complex \cite{tartakovsky2005modeling,hu2006multi}, and this delta function-based approach cannot discretize surface tension in an antisymmetric form, making it unable to ensure momentum conservation.
Following this, Lafaurie et al. \cite{lafaurie1994modelling} reformulated the surface tension in the CSF model as the divergence of a tensorial surface stress, thereby eliminating the need for curvature computation. Later, Hu and Adams \cite{hu2006multi} introduced this surface stress formulation into the SPH method and discretized its divergence (i.e., surface tension) in a momentum-conserving form.

When using the CSF model for simulations, numerical instabilities may arise at the fluid-fluid interface \cite{lafaurie1994modelling, rudman1998volume}. These instabilities can result in unrealistic interpenetrate or dispersion across the interface between the two fluids, potentially causing particle disorder at the fluid-fluid boundary \cite{morris2000simulating}. This, in turn, may lead to interface diffusion \cite{morris2000simulating}, compromising the accuracy of the simulation.
In previous research findings \cite{zhang2023improved,pozorski2024smoothed,hu2006multi}, it has been observed that particles at the interface exhibit a zigzag, sawtooth-like interlaced distribution, even within relatively short simulation times. As the simulation time extends, this behavior significantly undermines computational stability.

To address the long-standing issue of particle disorder at fluid-fluid interfaces, we conducted a detailed investigation and, for the first time, identified that this phenomenon is caused by a previously unrecognized numerical instability, which we define as \textit{zero-surface-energy modes}---zero-energy modes that specifically occur at fluid-fluid interfaces.
In SPH simulations of solid dynamics, zero-energy modes \cite{zhang2024essentially, ganzenmuller2015hourglass, zhang2024generalized, wu2024unified} are known to arise under specific particle distributions and velocity profiles, such as zigzag patterns. 
These modes are characterized by non-rigid body rotational deformations that fail to induce strain, resulting in inaccurate stress calculations. 
Similarly, in surface tension simulations, zigzag particle distributions pose significant challenges. 
Ideally, in regions of high curvature where surface tension is expected to be stronger, this zigzag distribution would naturally self-correct. 
However, the presence of zero-surface-energy modes results in an underestimation of surface tension in areas with a zigzag particle distribution, preventing the particle disorder from resolving.
By defining and identifying the underlying cause of these instabilities, this study makes a key contribution to the field. 
Specifically, we propose a novel penalty force method that compensates for the underestimated surface tension forces, effectively eliminating zigzag particle disorder. This method is formulated based on the use of surface stress \cite{hu2006multi, lafaurie1994modelling} for surface tension calculations, ensuring strict momentum conservation.
It is worth noting that the proposed penalty force term is also momentum-conserving and directly incorporated into the momentum equation without introducing additional computational complexity.

Another limitation commonly observed in previous SPH-based surface tension simulations is their restriction to relatively low Reynolds numbers ($Re < 50$) \cite{adami2010new,zhang2023improved,morris2000simulating,blank2023modeling,hu2006multi}. 
A recent SPH study \cite{marrone2025study} successfully increased the Reynolds number to 500, but simulations at higher Reynolds numbers have not yet been reported.
However, high Reynolds and Weber number cases, such as the collision of a liquid droplet with a solid surface at high speed, are common phenomena in natural processes and play a crucial role in a wide range of technological applications, including inkjet printing, pesticide application, and spray cooling.
In the multiphase flow framework developed in this study, we demonstrate the capability to simulate Reynolds numbers as high as 10000 and Weber numbers up to 25000. Even under such extreme conditions, our method remains stable and robust.
To the best of our knowledge, these are the highest Reynolds and Weber numbers ever reported in multiphase flow simulations employing particle-based methods.

The remainder of this article is organized as follows. Section \ref{governing-equation} introduces the governing equations for fluid dynamics. Section \ref{numerical-method} describes the multiphase Riemann-SPH framework employed in this study. 
Section \ref{surface-tension-formulation} presents a comprehensive description of the calculation of momentum-conserving surface tension and discusses the detection and mitigation of zero-surface-energy modes. 
Section \ref{numerical-examples} validates the stability and accuracy of the proposed method through several two-dimensional (2D) and three-dimensional (3D) benchmark cases and demonstrates its effectiveness under high Reynolds and Weber numbers. 
Finally, Section \ref{conclusions} concludes the study.
The computational codes used in this study have been open-sourced as part of the SPHinXsys project \cite{zhang2021sphinxsys}, available at \href{https://www.sphinxsys.org}{https://www.sphinxsys.org}.
%%%%%%%%%%%%%%%%%%%%%%%%%%%%%%%%%%%%%%%%%%%%%%%%%%%%%%%%%%%%%
%
% 2 Governing equations
%
%%%%%%%%%%%%%%%%%%%%%%%%%%%%%%%%%%%%%%%%%%%%%%%%%%%%%%%%%%%%%
\section{Governing equations}
\label{governing-equation}

The weakly compressible SPH (WCSPH) method is employed to model multiphase flows involving both viscosity and surface tension. 
In the Lagrangian framework, the governing equations are based on the conservation of mass and momentum, and can be expressed as
\begin{equation}
    \frac{\text{d} \rho }{\text{d} t} = -\mathbf \rho \nabla \cdot \mathbf v
   \label{continuity-equation}
\end{equation}
\begin{equation}
    \frac{\text{d} \mathbf v}{\text{d} t} = -\frac{1}{\mathbf \rho}\nabla \cdot p + \frac{1}{\mathbf \rho} \mathbf F^v + \frac{1}{\mathbf \rho} \mathbf F^s + \mathbf g
   \label{momentum-equation}
\end{equation}
where ${\rho}$ is the density, $\mathbf{v}$ is velocity, $p$ is the pressure, and $\mathbf{g}$ is the body force. 
$\mathbf F^v$ and $\mathbf F^s$ are viscous and surface tension forces, respectively.

To model incompressible flow under the weakly compressible assumption \cite{monaghan1994simulating, morris1997modeling}, an artificial isothermal equation of state is introduced to close Eq. \eqref{momentum-equation}.
\begin{equation}
    p=c_0^2(\rho - \rho_0)
   \label{EoS}
\end{equation}
where ${\rho_0}$ is the initial density and ${c_0}$ is the sound speed.
Typically, the density fluctuates by approximately 1\% \cite{morris1997modeling} when an artificial sound speed of $c_0 = 10U_{max}$ is applied, with $U_{max}$ being the maximum anticipated flow speed.
The speed of sound is consistently set to the same value for different phases of the fluid.

The viscous force $\mathbf F^v$ simplifies to its incompressible form.
\begin{equation}
    \mathbf F^v = \eta \nabla^2 \mathbf v
   \label{viscous-force}
\end{equation}
where $\eta$ is the dynamic viscosity.

Based on the CSF model \cite{brackbill1992continuum}, the surface force is represented as a volumetric force using the surface delta function $\delta_s$.
\begin{equation}
    \mathbf F^s = -\gamma \kappa  \mathbf n \delta_s
   \label{pre-surface-tension}
\end{equation}
where $\gamma$ is the surface tension coefficient at the interface of two different fluids, $\kappa$ is the curvature, $\mathbf n$ is the unit normal vector of the interface, and $\delta_s$ is the surface delta function.
This method for calculating surface tension does not conserve the total momentum within the system \cite{adami2010new}. 
The proposed momentum-conservative surface tension formulation will be introduced in detail in Section \ref{surface-tension-formulation}.
%%%%%%%%%%%%%%%%%%%%%%%%%%%%%%%%%%%%%%%%%%%%%%%%%%%%%%%%%%%%%
%
% 3 Numerical method
%
%%%%%%%%%%%%%%%%%%%%%%%%%%%%%%%%%%%%%%%%%%%%%%%%%%%%%%%%%%%%%
\section{Numerical method}
\label{numerical-method}
This section presents the numerical method employed in this study, including the SPH discretization of pressure and viscous forces, the transport-velocity formulation to mitigate tensile instability, and the time integration scheme.
The calculation of surface tension will be presented in Section \ref{surface-tension-formulation}.
%%%%%%%%%%%%%%%%%%%%%%%%%%%%%%%%%%%%%%%%%%%%%%%%%%%%%%%%%%%%%
% 3.1 SPH discretization
%%%%%%%%%%%%%%%%%%%%%%%%%%%%%%%%%%%%%%%%%%%%%%%%%%%%%%%%%%%%%
\subsection{Multiphase Riemann-SPH solver}
\label{SPH-discretization}

A low-dissipation Riemann solver \cite{zhang2017weakly, zhang2024riemann} is applied within the WCSPH framework to discretize the continuity equation and the pressure term in the momentum equation, as shown
\begin{equation}
    \frac{\text{d} \rho_i }{\text{d} t} = 2 \rho_i \sum_{j} (\mathbf v_{i}-\mathbf v^*) \dot {\nabla_i W_{ij}} V_j
   \label{continuity-equation-discrete}
\end{equation}
\begin{equation}
   \frac{\text{d} \mathbf v_i^p}{\text{d} t} = -2\frac{1}{\rho_i}\sum_{j} P^* {\nabla_i W_{ij}} V_j
  \label{normal-accelaration-discrete}
\end{equation}
where $W_{ij}=W({\mathbf r}_i- {\mathbf r}_j, h)$ is the kernel function. ${\mathbf r}$ and ${h}$ are the particle position and smoothing length, respectively. 
The subscripts ${i}$ and ${j}$ represents particle numbers, with $V$ being the particle volume.
The vector $\mathbf{e}_{ij}$ is a unit vector directed from particle ${j}$ to particle ${i}$. The term ${\nabla_i W_{ij}}=\frac{\partial  W({r}_{ij}, h)}{\partial {r}_{ij}} \mathbf e_{ij}$ denotes the gradient of the kernel function, where ${r}_{ij} = |\mathbf r_{i} - \mathbf r_{j}|$ is the distance between particles ${i}$ and ${j}$.
The superscript $p$ in $\frac{\text{d} \mathbf v_i^p}{\text{d} t}$ indicates the acceleration resulting from pressure.

The variables $\mathbf v^*$ and $P^*$, derived from the low-dissipation Riemann solver \cite{zhang2017weakly}, represent the solutions to an inter-particle Riemann problem in the direction of the unit vector that pointing from particle $i$ to $j$.
$\mathbf v^*$ and $P^*$ are defined as \cite{zhang2017weakly}
\begin{equation}
	\begin{cases}
		\mathbf v^* = U^*\mathbf e_{ij} + \left(\overline{\mathbf v}_{ij}-\overline{U}\mathbf e_{ij} \right) \vspace{0.1cm}\\
      \displaystyle P^* = \overline{p}_{ij} + \frac{\beta \rho_L c_L \rho_R c_R (U_L-U_R)}{\rho_L c_L + \rho_R c_R} 
	\end{cases}
   \label{v-p-Riemann}
\end{equation}
with
\begin{equation}
   U^* = \frac{\rho_L c_L U_L+\rho_R c_R U_R+P_L-P_R}{\rho_L c_L + \rho_R c_R}
  \label{U-Riemann}
\end{equation}
The subscripts $L$ and $R$ denote the left and right states from Riemann problem, respectively.
The left and right states are defined as
\begin{equation}
	\begin{cases}
		\left(\rho_L, U_L, P_L, c_L \right) = \left(\rho_i, - \mathbf v_i \cdot \mathbf e_{ij}, p_i, c_{0i} \right) \vspace{0.1cm}\\
		\left(\rho_R, U_R, P_R, c_R\right) = \left( \rho_j, - \mathbf v_j \cdot \mathbf e_{ij}, p_j, c_{0j} \right)
	\end{cases}
   \label{left-right-states}
\end{equation}
$\beta$ in Eq. \eqref{v-p-Riemann} is the dissipation limiter, which is defined as \cite{zhang2017weakly}
\begin{equation}
    \beta=\min \left\{
		3\max \left[
			\frac{(P_L+P_R)(U_L-U_R)}{\rho_L c_L+\rho_R c_R}, 0
		\right], 1.0
      \right\} 
   \label{dissipation-limiter}
\end{equation}

For multiphase flows, the particle-pair average values $\overline{\mathbf v}_{ij}$, $\overline{U}$, and $\overline{p}_{ij}$ are calculated as
\begin{equation}
	\begin{cases}
      \displaystyle \overline{U} = \frac{\rho_L U_L+\rho_R U_R}{\rho_L+\rho_R}\\
      \displaystyle \overline{\mathbf v}_{ij} = \frac{\rho_i\mathbf v_i+\rho_j\mathbf v_j}{\rho_i+\rho_j}\\
      \displaystyle \overline{p}_{ij} = \frac{\rho_i p_j + \rho_j p_i}{\rho_i+\rho_j}
	\end{cases}
   \label{particle-pair-average}
\end{equation}
The viscous force $\mathbf F^v$ is discretized as \cite{hu2006multi,morris1997modeling}
\begin{equation}
    \mathbf F^v_i = 2 \frac{1}{\rho_i} \sum_{j} \frac{2\eta_i \eta _j}{\eta_i + \eta _j} \frac{\mathbf v_{ij}}{r_{ij}} \frac{\partial W_{ij}}{\partial {r}_{ij}} V_j
   \label{viscous-force-discrete}
\end{equation}
where $\mathbf v_{ij}=\mathbf v_{i}-\mathbf v_{j}$.
%%%%%%%%%%%%%%%%%%%%%%%%%%%%%%%%%%%%%%%%%%%%%%%%%%%%%%%%%%%%%
% 3.2 Transport-velocity formulation
%%%%%%%%%%%%%%%%%%%%%%%%%%%%%%%%%%%%%%%%%%%%%%%%%%%%%%%%%%%%%
\subsection{Transport-velocity formulation}
\label{transport-velocity}
SPH often exhibits tensile instability \cite{zhang2024essentially,swegle1995smoothed,gray2001sph}, characterized by particle clustering and the formation of non-physical voids. 
In this study, the transport-velocity formulation \cite{adami2013transport} is employed as a numerical technique to prevent the occurrence of tensile instability. 
The latest improvement to the transport-velocity formulation \cite{zhang2025towards, zhang2025unified} involves correcting particle positions within their support domains to achieve local zeroth-order consistency, thereby ensuring a more uniform particle distribution and effectively eliminating tensile instability.
An advection term $\Delta \mathbf r$, determined by  local particle consistency, is used to correct the particle position, which is defined as
\begin{equation}
   \Delta \mathbf r_i =\alpha  h^2 \sum_{j} (\mathbf B_i + \mathbf B_j) \nabla_i W_{ij} V_j
  \label{consistency-term-Bo}
\end{equation}
where $\alpha$ is a coefficient and can be generally set to 0.2 \citep{zhang2025towards}.
$\mathbf B_i$ is the correction matrix for the kernel gradient, and is defined as \citep{randles1996smoothed}
\begin{equation}
    \mathbf B_i = - \left({\sum_{j} \mathbf r_{ij} \otimes {\nabla_i W_{ij}}  V_j} \right)^{-1}
   \label{correction-matrix}
\end{equation}
where $\mathbf r_{ij}=\mathbf r_{i}-\mathbf r_{j}$.
%%%%%%%%%%%%%%%%%%%%%%%%%%%%%%%%%%%%%%%%%%%%%%%%%%%%%%%%%%%%%
% 3.3 Time-step criteria
%%%%%%%%%%%%%%%%%%%%%%%%%%%%%%%%%%%%%%%%%%%%%%%%%%%%%%%%%%%%%
\subsection{Time-step criteria}
\label{time-step}

To enhance computational efficiency, a dual-criteria time stepping \cite{zhang2020dual} approach is adopted for iterative updates. 
This approach consists of an advection time step $\bigtriangleup t_{ad}$ and an acoustic time step $\bigtriangleup t_{ac}$. 
During the advection time step, the primary focus is on updating the particle configuration, including the particle-neighbor list as well as kernel values and gradients. 
In contrast, the acoustic time step mainly handles the time integration of the governing equations. 
Typically, the advection time step is larger than the acoustic time step, allowing for improved computational efficiency by reducing the frequency of particle configuration updates.

In this study, the computations of viscous forces and the advection term $\Delta \mathbf r_i$ in the transport-velocity formulation are performed during the advection time step. While all other physical quantities, including surface tension, are computed during the acoustic time step, due to surface tension calculations require smaller time steps.

If there are two fluids, denoted as $k$ and $l$, the advection time step for any given fluid $k$ is expressed as
\begin{equation}
   \bigtriangleup t_{ad}^k = 0.1\min \left [ \frac{h^k}{\left\lvert  \mathbf v^k \right\rvert}_{max}, \frac{\rho_0^k {(h^k)}^2}{\eta^k} \right ]
  \label{advection-time-step}
\end{equation}
The acoustic time step for fluid $k$ is defined as
\begin{equation}
   \bigtriangleup t_{ac}^k = 0.6\min \left \{ \frac{h^k}{c_0^k+\left\lvert  \mathbf v^k \right\rvert}_{max}, {\left [ \frac{\rho_0^k {(h^k)}^3}{2\pi \gamma} \right ]}^{0.5} \right \}
  \label{acoustic-time-step}
\end{equation}
where $\mathbf v^k_{max}$ is the maximum particle advection speed, $\eta^k$ is the dynamic viscosity, $c_0^k$ is the sound speed, and $\gamma$ is the surface tension coefficient between fluids $k$ and $l$.

Then the time step for the entire system is determined by the minimum of the time steps for all fluids.
\begin{equation}
   \begin{cases}
   \bigtriangleup t_{ad} = \min \left ( \bigtriangleup t_{ad}^k, \bigtriangleup t_{ad}^l \right ) \\
   \bigtriangleup t_{ac} = \min \left ( \bigtriangleup t_{ac}^k, \bigtriangleup t_{ac}^l \right )
\end{cases}
  \label{time-step-size}
\end{equation}

%%%%%%%%%%%%%%%%%%%%%%%%%%%%%%%%%%%%%%%%%%%%%%%%%%%%%%%%%%%%%
%
% 4 Momentum-conservative surface tension formulation
%
%%%%%%%%%%%%%%%%%%%%%%%%%%%%%%%%%%%%%%%%%%%%%%%%%%%%%%%%%%%%%
\section{Momentum-conservative surface tension formulation}
\label{surface-tension-formulation}
%%%%%%%%%%%%%%%%%%%%%%%%%%%%%%%%%%%%%%%%%%%%%%%%%%%%%%%%%%%%%
% 4.1 Surface tension formulation
%%%%%%%%%%%%%%%%%%%%%%%%%%%%%%%%%%%%%%%%%%%%%%%%%%%%%%%%%%%%%
\subsection{Surface tension formulation}
\label{surface-tension}
\begin{figure}[htb!]
	\centering
	\includegraphics[trim = 0cm 0cm 0cm 0cm, clip,width=.5\textwidth]{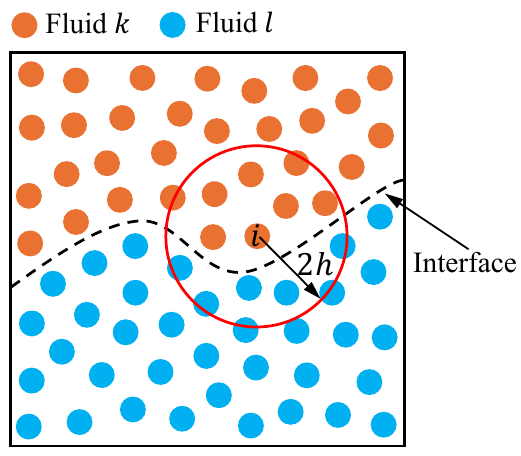}
	\caption{Illustration of two fluid phases $k$ and $l$.}
	\label{figs:illustration_fluid_kl}
\end{figure}

Unlike previous methods that calculate surface tension based on the surface curvature \cite{adami2010new, brackbill1992continuum}, this study proposes a momentum-conserving surface tension formulation based on surface stress \cite{hu2006multi, lafaurie1994modelling}.
In contrast to the study of Hu and Adami \cite{hu2006multi}, which derives surface tension solely from the contacting fluid, the present approach incorporates contributions from both the fluid itself and the adjacent fluid.
Firstly, the calculation of surface stress is defined as \cite{hu2006multi,lafaurie1994modelling}
\begin{equation}
    \mathbf \Pi  = \gamma \left (\mathbf I - \hat{\mathbf n}  \hat{\mathbf n}^T \right ) \left\lvert \nabla C \right\rvert 
   \label{surface-stress}
\end{equation}
And the surface tension force $\mathbf F^s$ is then obtained by \cite{hu2006multi,lafaurie1994modelling}
\begin{equation}
    \mathbf F^s = \nabla \cdot \mathbf \Pi
   \label{surface-force}
\end{equation}
Here, $\nabla C$ is the gradient of the color function $C$, and color function is used to identify the interface between two fluids \cite{brackbill1992continuum}.
$\hat{\mathbf n}$ is the normalized interface direction, which is defined as
\begin{equation}
   \hat{\mathbf n} = \frac{\nabla C}{\left\lvert \nabla C \right\rvert}
   \label{interface-normal}
\end{equation}
If particle $i$ is the target particle for computation, as shown in Fig. \ref{figs:illustration_fluid_kl}, its color function value is defined as $C_i = 1$. 
Assuming particle $i$ belongs to fluid phase $k$, i.e., $i\in k$, the color function value for its neighboring particle $j$ is $C_j = 1$ if $j\in k$. Otherwise, if $j$ belongs to the other phase $l$, the color function value is $C_j = 0$.

The gradient of the color function is calculated as
\begin{equation}
   \nabla C_i = \sum_{j} 2\varphi_i C_{ij} \nabla_i W_{ij} V_j
   \label{color-gradient}
\end{equation}
Here, $C_{ij}=C_{i}-C_{j}$. 
Clearly, $C_{ij}=0$ when $i$ and $j$ belong to the same phase, and $C_{ij}=1$ when $i$ and $j$ belong to different phases.
Therefore, in practice, the color gradient is only calculated at the interfaces between different phases.
$\varphi_i$ represents the ratio of the density of particle $i$ to the total density of particles $i$ and $j$, and is defined as
\begin{equation}
   \varphi_i = \frac{\rho_{0,i}}{\rho_{0,i}+\rho_{0,j}}, \
   \varphi_j = \frac{\rho_{0,j}}{\rho_{0,i}+\rho_{0,j}}
   \label{density-ratio}
\end{equation} 
When $i$ and $j$ belong to the same fluid phase, $2\varphi_i=2\varphi_j=1$.

The surface stress $\mathbf \Pi$ for particle $i$ is then calculated as
\begin{equation}
   \mathbf \Pi_i = \gamma \left (\mathbf I - \hat{\mathbf n}_i  \hat{\mathbf n}_i^T \right ) \left\lvert \nabla C_i \right\rvert 
  \label{surface-stress-i}
\end{equation}
Considering that the neighboring particles of particle $i$ originate from two different fluid phases, $k$ and $l$, the surface tension force is calculated as
\begin{equation}
   \mathbf F^s_i = \mathbf F^{s,k}_{i} + \mathbf F^{s,l}_{i}
   \label{surface-force-i}
\end{equation}
where $\mathbf F^{s,k}_{i}$ and $\mathbf F^{s,l}_{i}$ are the surface tension forces acting on particle $i \ (i\in k)$ from fluid phases $k$ and $l$, respectively.
Based on Eq. \eqref{surface-force}, $\mathbf F^{s,k}_{i}$ and $\mathbf F^{s,l}_{i}$ are calculated as
\begin{equation}
   \begin{cases}
      \displaystyle \mathbf F^{s,k}_{i} = \frac{m_i}{\rho_i} \sum_{j \in k} \left ( \mathbf \Pi_{i} + \mathbf \Pi_{j} \right ) \nabla_i W_{ij} V_j \\
      \displaystyle \mathbf F^{s,l}_{i} = \frac{m_i}{\rho_i} \sum_{j \in l} \left ( 2\varphi_j \mathbf \Pi_{i} + 2\varphi_i \mathbf \Pi_{j} \right ) \nabla_i W_{ij} V_j
   \end{cases}
   \label{surface-force-kl}
\end{equation}
%%%%%%%%%%%%%%%%%%%%%%%%%%%%%%%%%%%%%%%%%%%%%%%%%%%%%%%%%%%%%
% 4.2 Zero-surface-energy modes
%%%%%%%%%%%%%%%%%%%%%%%%%%%%%%%%%%%%%%%%%%%%%%%%%%%%%%%%%%%%%
\subsection{Zero-surface-energy modes}
\label{zero-surface-energy}
In surface tension computations, specific particle distributions can result in the underestimation of surface stress at the interface. 
This underestimation is primarily caused by errors in the estimation of the color gradient. 

\begin{figure}[htb!]
	\centering
	\includegraphics[trim = 0cm 0cm 0cm 0cm, clip,width=1\textwidth]{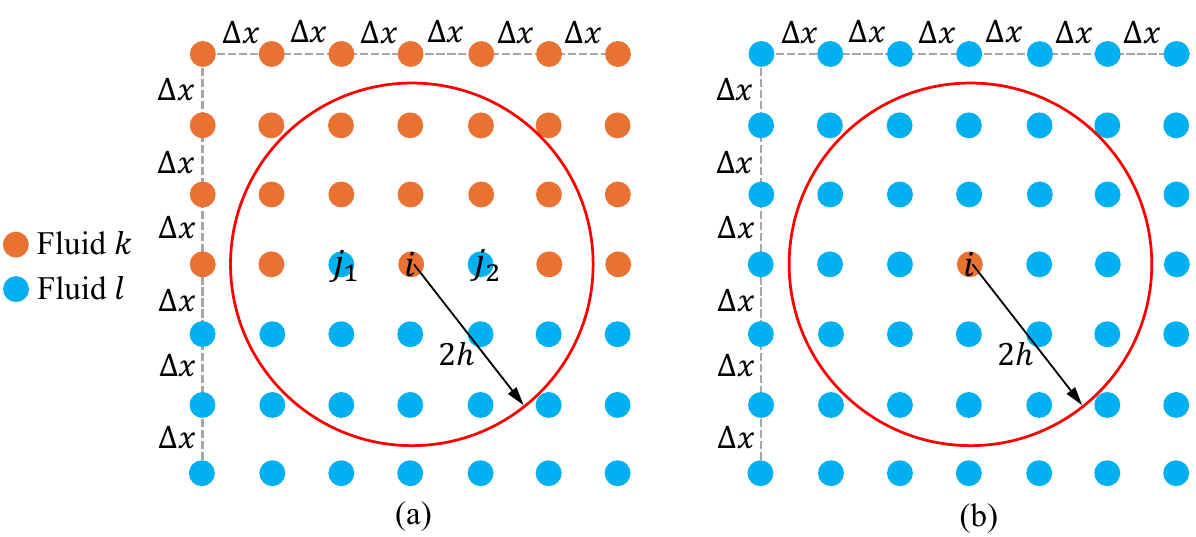}
	\caption{Illustration of zero-surface-energy modes: (a) a general case; (b) an extreme case.}
	\label{figs:zero_surface_energy_modes}
\end{figure}

Fig. \ref{figs:zero_surface_energy_modes} illustrates two representative cases of zero-surface-energy modes. 
Fig. \ref{figs:zero_surface_energy_modes}a shows a common scenario. 
When calculating the color gradient of particle $i$ based on Eq. \eqref{color-gradient}, the contributions of particles $j_1$ and $j_2$ to the color gradient, denoted as $\nabla C_{i-j_1}$ and $\nabla C_{i-j_2}$, cancel each other out ($\nabla C_{i-j_1} + \nabla C_{i-j_2} = 0$). 
This mutual cancellation leads to an underestimation of the color gradient. 
Fig. \ref{figs:zero_surface_energy_modes}b presents an extreme case: for any pair of neighbor particles $j$ centered around particle $i$, their contributions to the color gradient completely cancel each other out, similar to Fig. \ref{figs:zero_surface_energy_modes}a. 
As a result, the final color gradient for particle $i$ becomes zero ($\nabla C_i=0$). 
According to Eq. \eqref{surface-stress-i}-Eq. \eqref{surface-force-kl},this computational error in the color gradient will further lead to the underestimation of the surface tension force. 

Drawing on the treatment of zero-energy modes in solid materials within the framework of updated Lagrangian SPH \cite{zhang2024generalized}, this paper introduces a penalty force $\hat{\mathbf f}_{ij}$ to compensate for the erroneously estimated surface tension force between each particle pair $i$ and $j$.
Firstly, assuming that the color function is locally linear, we can obtain the difference in color function values, $C_{ij}$, for each pair of particles $i$ and $j$.
\begin{equation}
   {C}_{ij}^{linear}=\frac{1}{2}\left(\nabla C_{i} + \nabla C_{j}\right)\cdot \mathbf r_{ij}
  \label{Cij_linear}
\end{equation}
Then the difference, $\widehat{C}_{ij}$, between the actual color function difference ${C_{ij}}$ and the linearly predicted color function difference ${C}_{ij}^{linear}$ is calculated as
\begin{equation}
   \widehat{C}_{ij} = C_{ij} - {C}_{ij}^{linear} = C_{ij} - \overline{\nabla C}_{ij} \cdot \mathbf r_{ij}
  \label{Cij_hat}
\end{equation}
where $\overline{\nabla C}_{ij} = \frac{1}{2}(\nabla C_{i} + \nabla C_{j})$.

The penalty force $\hat{\mathbf f}_{ij}$ is then defined as
\begin{equation}
   \hat{\mathbf f}_{ij} = \frac{m_i}{\rho_i} \sum_{j} \xi \gamma \frac{\widehat{C}_{ij}}{r_{ij}} \nabla_i W_{ij} V_j
  \label{penalty-force}
\end{equation}
where $\xi$ is a coefficient, which is used to control the magnitude of the penalty force.

It can be observed that the correction term $\hat{\mathbf f}_{ij}$ does not satisfy the principle of momentum conservation (i.e., $\hat{\mathbf f}_{ij} + \hat{\mathbf f}_{ji}\neq 0$).
To address this issue, we further modified $\widehat{C}_{ij}$ and reformulated it as a tensor $\widehat{\mathbf C}_{ij}$ to ensure consistency with momentum conservation.
$\widehat{\mathbf C}_{ij}$ is defined as
\begin{equation}
   \widehat{\mathbf C}_{ij} = C_{ij} \mathbf I - \frac{\overline{\nabla C}_{ij} \otimes \mathbf r_{ij}}{\left\lvert \overline{\nabla C}_{ij} \otimes \mathbf r_{ij} \right\rvert } \otimes \left ( \overline{\nabla C}_{ij} \otimes \mathbf r_{ij} \right )
  \label{Cij_hat_tensor}
\end{equation}
The penalty force $\hat{\mathbf f}_{ij}$ is reformulated as
\begin{equation}
   \hat{\mathbf f}_{ij} = \frac{m_i}{\rho_i} \sum_{j} \xi \gamma \frac{\widehat{\mathbf C}_{ij}}{r_{ij}} \nabla_i W_{ij} V_j
  \label{penalty-force-tensor}
\end{equation}
Then the surface tension force acting on particle $i \ (i\in k)$, as shown in Eq. \eqref{surface-force-kl}, can be improved by adding the penalty force $\hat{\mathbf f}_{ij}$.
\begin{equation}
   \begin{cases}
      \displaystyle \mathbf F^{s,k}_{i} = \frac{m_i}{\rho_i} \sum_{j \in k} \left ( \mathbf \Pi_{i} + \mathbf \Pi_{j} 
      +\xi \gamma \frac{\widehat{\mathbf C}_{ij}}{r_{ij}}
      \right ) \nabla_i W_{ij} V_j \\
      \displaystyle \mathbf F^{s,l}_{i} = \frac{m_i}{\rho_i} \sum_{j \in l} \left ( 2\varphi_j \mathbf \Pi_{i} + 2\varphi_i \mathbf \Pi_{j} 
      - 4 \varphi_i \varphi_j \xi \gamma \frac{\widehat{\mathbf C}_{ij}}{r_{ij}}\right ) \nabla_i W_{ij} V_j
   \end{cases}
   \label{surface-force-kl-HC}
\end{equation}
The last term in the parentheses on the right-hand side of Eq. \eqref{surface-force-kl-HC} represents the penalty force correction term. 
Within the computational framework of this study, it is noted that the interfacial tension within a phase acts tangentially to the interface, while the interfacial tension between phases acts normally to the interface. 
Accordingly, the penalty forces within and between phases are projected onto their respective directions, yielding
\begin{equation}
	\begin{cases}
	   \displaystyle \mathbf F^{s,k}_{i} = \frac{m_i}{\rho_i} \sum_{j \in k} \left ( \mathbf \Pi_{i} + \mathbf \Pi_{j} 
	   +\xi \gamma \overline{\mathbf T}_{ij}  \frac{\widehat{\mathbf C}_{ij}}{r_{ij}}
	   \right ) \nabla_i W_{ij} V_j \\
	   \displaystyle \mathbf F^{s,l}_{i} = \frac{m_i}{\rho_i} \sum_{j \in l} \left ( 2\varphi_j \mathbf \Pi_{i} + 2\varphi_i \mathbf \Pi_{j} 
	   -4 \varphi_i \varphi_j \xi \gamma \overline{\mathbf N}_{ij} \frac{\widehat{\mathbf C}_{ij}}{r_{ij}}\right ) \nabla_i W_{ij} V_j
	\end{cases}
	\label{surface-force-kl-final}
\end{equation}
Here, $i \in k$. $\overline{\mathbf T}_{ij}=0.5({\mathbf T}_{i}+{\mathbf T}_{j})$ and ${\mathbf T}_{i}=\mathbf I - \hat{\mathbf n}_i  \hat{\mathbf n}_i^T$.
$\overline{\mathbf N}_{ij}=0.5({\mathbf N}_{i}+{\mathbf N}_{j})$ and ${\mathbf N}_{i}=\hat{\mathbf n}_i  \hat{\mathbf n}_i^T$.

In this study, the value of the parameter $\xi$ was determined through numerical experiments. 
For two fluids with a density ratio of $10^n$, $\xi$ is given by the relation $\xi = 1.3n + 0.6$. 
This indicates that the parameter depends solely on the density ratio and requires no further tuning.
%%%%%%%%%%%%%%%%%%%%%%%%%%%%%%%%%%%%%%%%%%%%%%%%%%%%%%%%%%%%%
%
% 5 Numerical examples
%
%%%%%%%%%%%%%%%%%%%%%%%%%%%%%%%%%%%%%%%%%%%%%%%%%%%%%%%%%%%%%
\section{Numerical examples}
\label{numerical-examples}
This section first validates the stability and accuracy of the proposed multiphase framework in simulating surface tension through several well-established benchmark cases with theoretical solutions, including the oscillating drop, drop deformation in shear flow, and square droplet oscillation. 
Subsequently, the high-velocity droplet impact case is used to demonstrate that the framework can produce stable and physically reasonable results even under conditions of high Reynolds and Weber numbers.
The 5th-order Wendland kernel \cite{wendland1995piecewise}, with a smoothing length of $h = 1.3dp$ and a cut-off radius of $2.6dp$, is applied to all cases, where $dp$ represents the initial particle spacing.
%%%%%%%%%%%%%%%%%%%%%%%%%%%%%%%%%%%%%%%%%%%%%%%%%%%%%%%%%%%%%
% 5.1 2D oscillating drop
%%%%%%%%%%%%%%%%%%%%%%%%%%%%%%%%%%%%%%%%%%%%%%%%%%%%%%%%%%%%%
\subsection{Oscillating drop}
\label{oscillating-drop}
\begin{figure}[htb!]
	\centering
	\includegraphics[trim = 0cm 0cm 0cm 0cm, clip,width=0.5\textwidth]{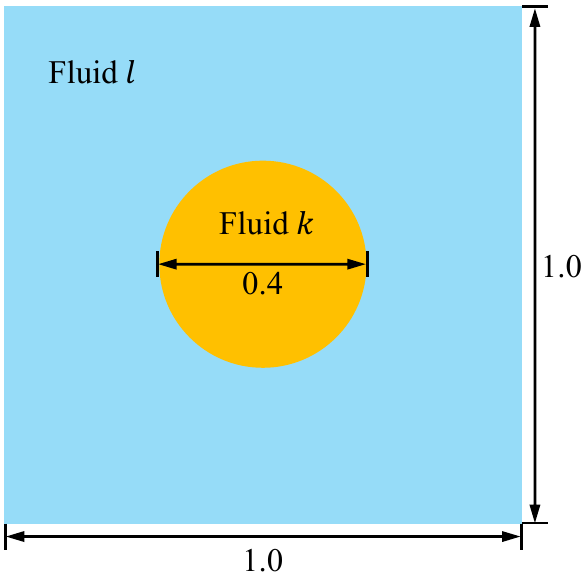}
	\caption{Oscillating drop: model setup.}
	\label{figs:oscillating_drop_setup}
\end{figure}

Firstly, the oscillation of a circular liquid-droplet under the action of capillary forces is simulated, to validate the proposed momentum-conservative surface tension formulation.
As shown in Fig. \ref{figs:oscillating_drop_setup}, the size of computational domain is $1.0 \times 1.0$, and the radius of the droplet is $R=0.2$ \cite{adami2010new}.
The drop is located at the center of the computational domain with the following prescribed initial velocity.
\begin{equation}
   \begin{cases}
      \displaystyle v_x = V_0 \frac{x}{r_0}\left ( 1-\frac{y^2}{r_0r} \right ) \text{exp}\left (-\frac{r}{r_0} \right ) \\
      \displaystyle v_y = - V_0 \frac{y}{r_0}\left ( 1-\frac{x^2}{r_0r} \right ) \text{exp}\left (-\frac{r}{r_0}\right )
   \end{cases}
   \label{initial-velocity}
\end{equation}
The velocity field is divergence-free, resulting in the initial fluid cylinder being stretched along the x-axis when $V_0>0$.
In this case, $V_0=10$, $r_0=0.05$, $\rho_k=1$, $\eta_k=0.05$, and the surface tension coefficient $\gamma=1.0$.

\begin{figure}[htb!]
	\centering
	\includegraphics[trim = 0cm 0cm 0cm 0cm, clip,width=1\textwidth]{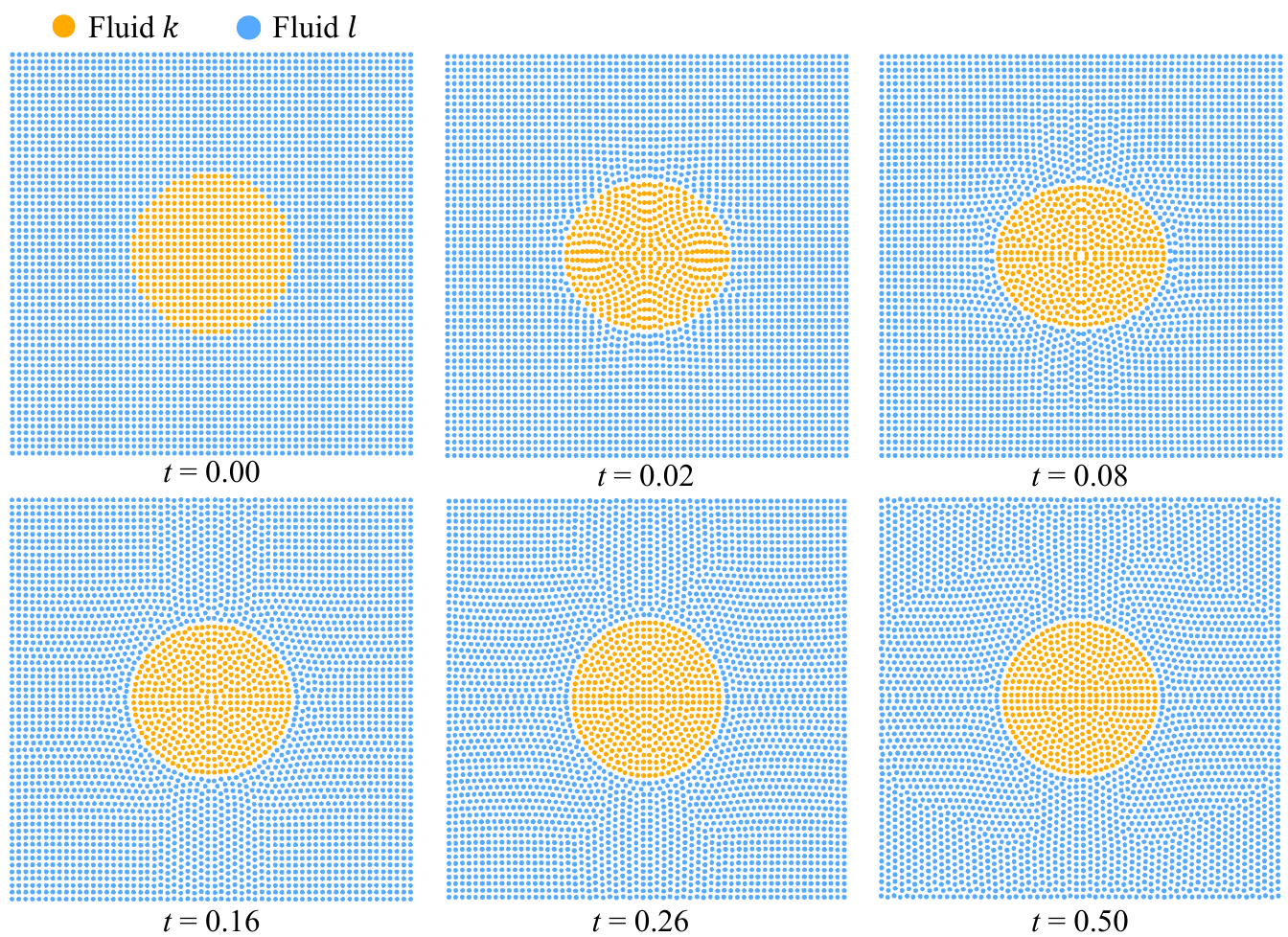}
	\caption{Oscillating drop: positions of droplet particles at $t=0.00$, 0.02, 0.08, 0.16, 0.26 and 0.50. Here, $\rho_k / \rho_l=1$ and $R/dp=12$.}
	\label{figs:oscillating_drop_D1_snapshot}
\end{figure}

\begin{figure}[htb!]
	\centering
	\includegraphics[trim = 0cm 0cm 0cm 0cm, clip,width=0.5\textwidth]{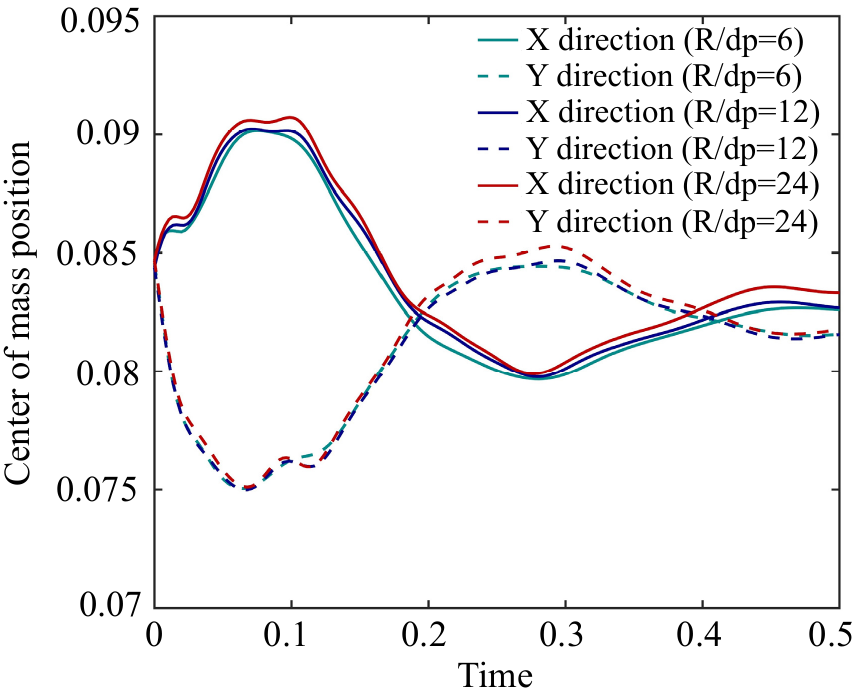}
	\caption{Oscillating drop: convergence of center of mass position. Here, $\rho_k / \rho_l=1$.}
	\label{figs:oscillating_drop_D1_convergence}
\end{figure}

For the first scenario, the density and viscosity of phase $l$ and phase $k$ are identical, with $\rho_l=1$, $\eta_l=0.05$. Fig. \ref{figs:oscillating_drop_D1_snapshot} illustrates the positions of the droplet particles at different time steps. 
Initially, the particles inside the droplet are distributed uniformly on the lattice, and over time, they evolve into a smooth interfacial distribution. 
Compared to previous studies, the droplet's deformation at the same time step demonstrates strong consistency \cite{adami2010new, hu2006multi}.
Fig. \ref{figs:oscillating_drop_D1_convergence} presents the position of the mass center of the particles located in the upper right-quarter section of the droplet. 
To verify convergence, results for three different resolutions are provided: $H/dp=6$, 12, and 24, corresponding to 900, 3600, and 14,400 particles, respectively. 
These resolutions are consistent with those used in the literature \cite{adami2010new, hu2006multi}.

For the second case, the physical parameters of phase $k$ remain unchanged, while the density and viscosity of phase $l$ are set to $\rho_l=1\times 10^{-3}$ and $\eta_l=5\times 10^{-4}$, respectively. 
This results in a density ratio of 1000 and a viscosity ratio of 100 between phase $k$ and phase $l$. Under the assumption of small deformation, the theoretical oscillation period of the droplet is given as 
\begin{equation}
   \tau  = 2\pi \sqrt{\frac{R^3 \rho_k}{6\gamma}}
  \label{theoretical-period}
\end{equation}
To validate the accuracy of the proposed method against the theoretical value, a small initial velocity $V_0=1$ is introduced.

\begin{figure}[htb!]
	\centering
	\includegraphics[trim = 0cm 0cm 0cm 0cm, clip,width=1\textwidth]{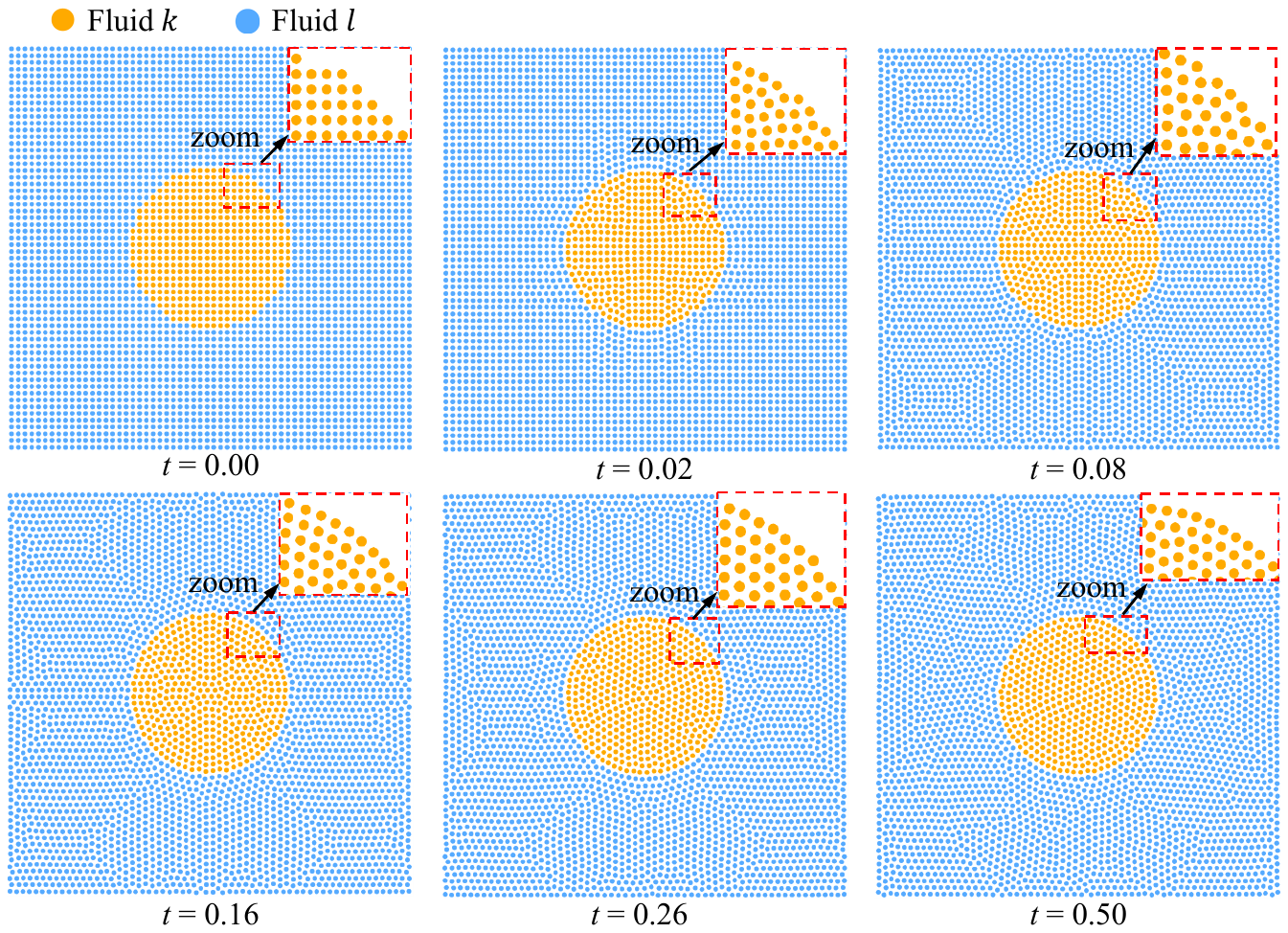}
	\caption{Oscillating drop: positions of droplet particles at $t=0.00$, 0.02, 0.08, 0.16, 0.26 and 0.50. Here, $\rho_k / \rho_l=1000$ and $R/dp=12$.}
	\label{figs:oscillating_drop_D1000_snapshot}
\end{figure}

\begin{figure}[htb!]
	\centering
	\includegraphics[trim = 0cm 0cm 0cm 0cm, clip,width=0.5\textwidth]{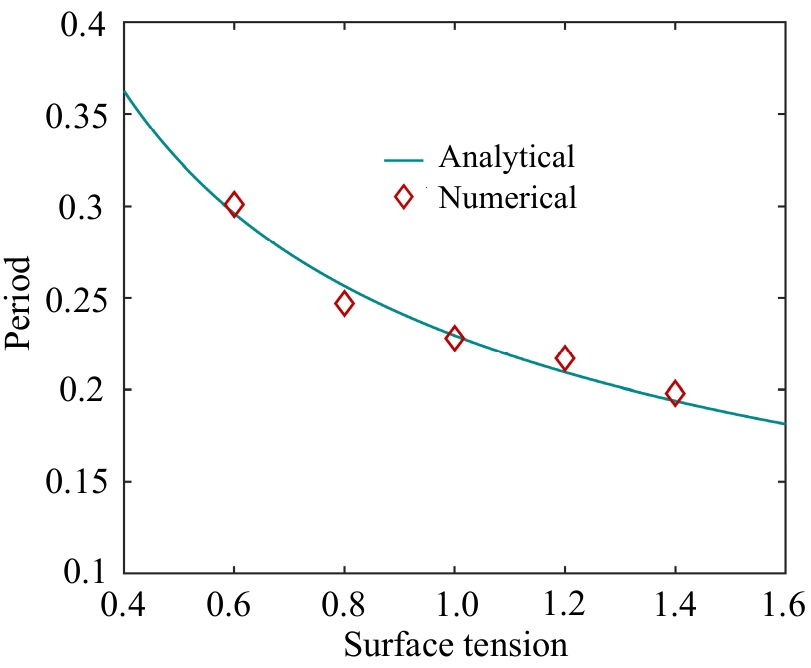}
	\caption{Oscillating drop: comparison of oscillation period between the numerical and analytic results. Here, $\rho_k / \rho_l=1000$.}
	\label{figs:oscillating_drop_D1000_period}
\end{figure}

Fig. \ref{figs:oscillating_drop_D1000_snapshot} shows the particle distribution at different time steps for the case with a density ratio of 1000. 
Due to the relatively small initial velocity, the changes in the particle contour are not significant. Therefore, an enlarged inset image is included in the top-right corner to illustrate the particle distribution at the droplet surface. 
Under such low initial velocity, surface tension dominates the droplet's evolution, eventually resulting in a circular shape with particles uniformly distributed along the interface.
Fig. \ref{figs:oscillating_drop_D1000_period} compares the theoretical values of the oscillation period with the numerical results under different surface tension values. 
The results demonstrate good agreement between the theoretical and numerical values, with a maximum error of less than 5\%.

\begin{figure}[htb!]
	\centering
	\includegraphics[trim = 0cm 0cm 0cm 0cm, clip,width=1\textwidth]{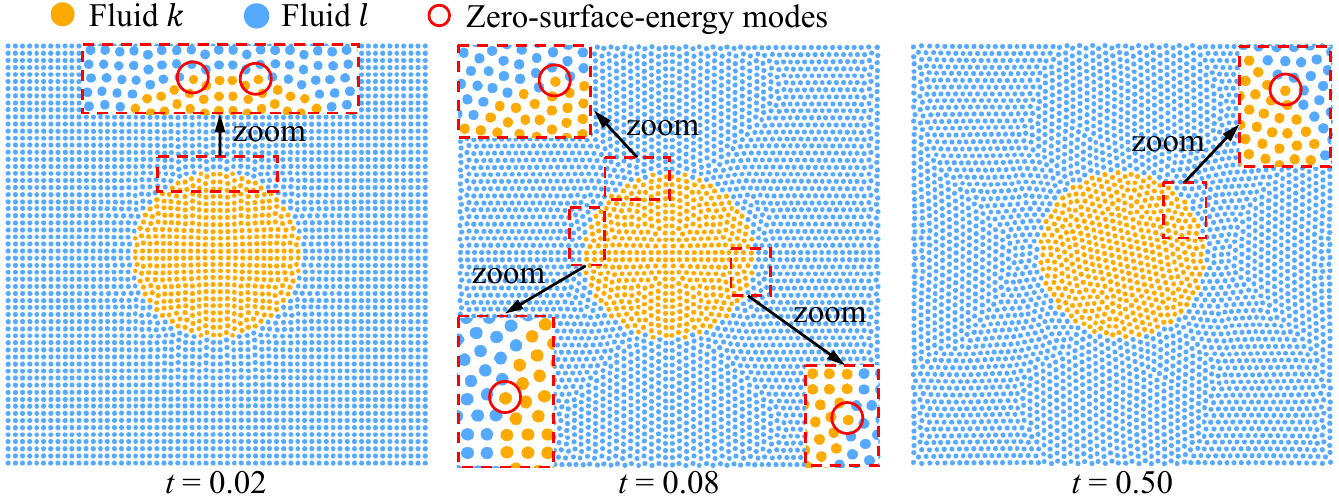}
	\caption{Oscillating drop: illustration of zero-surface-energy modes occurring when no penalty force is applied. $\rho_k / \rho_l=1000$ and $R/dp=12$.}
	\label{figs:oscillating_drop_D1000_HC0}
\end{figure}

In Section~\ref{zero-surface-energy}, we introduced the penalty force to eliminate zero-surface-energy modes. Here, we will demonstrate the results when the penalty force is not applied. 
As shown in Fig.~\ref{figs:oscillating_drop_D1000_HC0}, the particle distributions at different time steps are presented for the case without the penalty force.
Some enlarged local views are added to provide a clearer depiction of the particle distribution at the interface between the two fluids.
It can be observed that at the interface, particles belonging to different fluids interpenetrate, forming a jagged pattern, which corresponds to the zero-surface-energy modes illustrated in Fig.~\ref{figs:zero_surface_energy_modes}.
Theoretically, regions with sharp corners should experience higher surface tension, which would drive the interface to evolve toward a smoother configuration. 
However, due to the presence of zero-surface-energy modes, the surface tension acting on the particles is underestimated, as analyzed in Section~\ref{zero-surface-energy}. Consequently, the sharp corners fail to smooth out as they should.
%%%%%%%%%%%%%%%%%%%%%%%%%%%%%%%%%%%%%%%%%%%%%%%%%%%%%%%%%%%%%
% 5.2 drop deformation in shear flow
%%%%%%%%%%%%%%%%%%%%%%%%%%%%%%%%%%%%%%%%%%%%%%%%%%%%%%%%%%%%%
\subsection{Drop deformation in shear flow}
\label{drop-shear}
The deformation of a liquid droplet in a Couette flow \cite{adami2010new, hu2006multi} is simulated to further validate the proposed surface tension formulation.
The computational domain is a square with a size of $L_x=L_y=8.0$, and the droplet with radius $R=1$ is initially located at the center of the domain.
The initial particle spacing is set to $R/12$, resulting in a total of 9216 particles within the computational domain, consistent with the configuration in previous SPH studies. \cite{adami2010new, hu2006multi}. 
Velocities of $\pm U_{wall}$ are applied to the upper and lower no-slip wall boundaries, while periodic boundary conditions are imposed on the left and right sides of the computational domain.
The density and viscosity of phase $l$ are identical to those of phase $k$ \cite{hu2006multi}, i.e., $\rho_l=\rho_k$, $\eta_l=\eta_k$.
The droplet undergoes deformation in the flow and eventually reaches an equilibrium state under the competing effects of viscous forces and surface tension. The ratio between these two forces is characterized by the capillary number. If the shear rate is denoted as $G=2U_{wall}/L_y$, the capillary number ($Ca$) and Reynolds number ($Re$) are expressed as \cite{adami2010new}
\begin{equation}
   Ca = \frac{G \eta_k R}{\gamma}, \quad Re = \frac{\rho_k G R^2}{\eta_k}
   \label{capillary-Re}
\end{equation}
When the droplet reaches a stable state, it assumes an elliptical shape. The degree of deformation is characterized by the deformation parameter $D=(a-b)/(a+b)$, where $a$ and $b$ represent the major and minor axes of the final elliptical shape, respectively.
\begin{figure}[htb!]
	\centering
	\includegraphics[trim = 0cm 0cm 0cm 0cm, clip,width=0.5\textwidth]{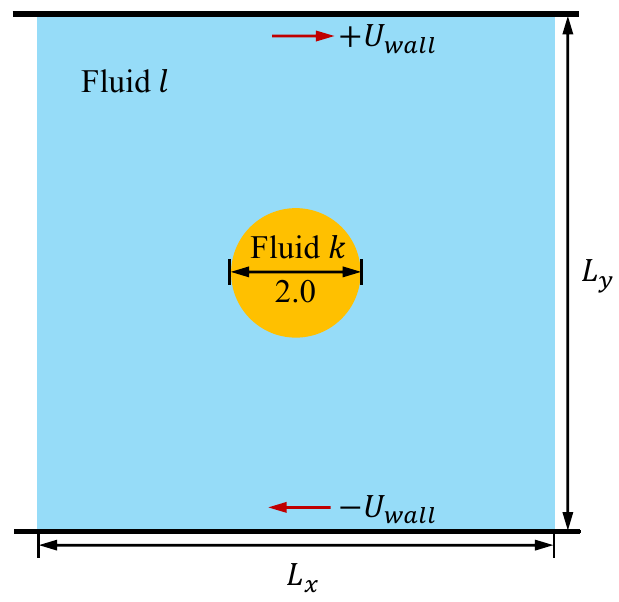}
	\caption{2D drop deformation in shear flow: model setup.}
	\label{figs:2d_drop_shear_setup}
\end{figure}

\begin{figure}[htb!]
	\centering
	\includegraphics[trim = 0cm 0cm 0cm 0cm, clip,width=1\textwidth]{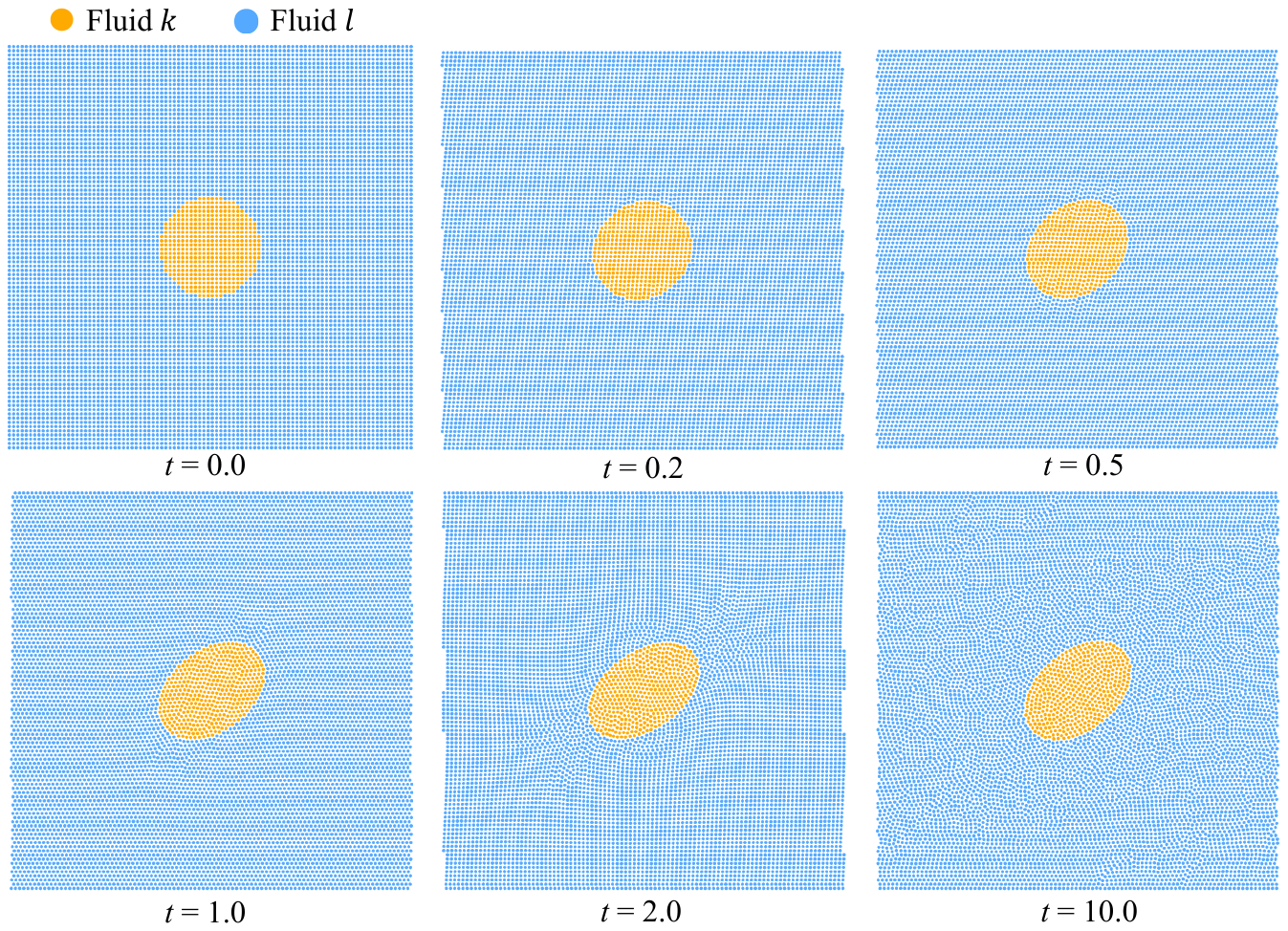}
	\caption{2D drop deformation in shear flow: positions of droplet particles and shearing fluid particles at $t=0.0$, 0.2, 0.5, 1.0, 2.0 and 10.0. Here, $Ca=0.2$, $Re=1.0$, and $R/dp=12$.}
	\label{figs:2d_drop_shear_snapshot}
\end{figure}

\begin{figure}[htb!]
	\centering
	\includegraphics[trim = 0cm 0cm 0cm 0cm, clip,width=0.5\textwidth]{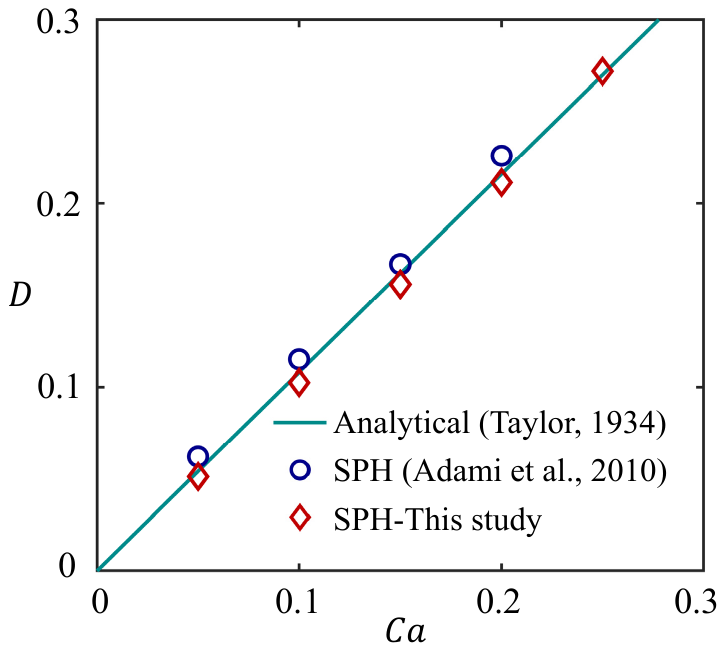}
	\caption{2D drop deformation in shear flow: comparison of the deformation parameter $D$ as a function of the capillary number $Ca$
   among the theoretical solution \cite{taylor1934formation}, Adami et al.'s SPH results \cite{adami2010new}, and the present SPH results. Here, $R/dp=12$.}
	\label{figs:2d_drop_shear_D}
\end{figure}

Fig. \ref{figs:2d_drop_shear_snapshot} shows the positions of the droplet particles and shearing fluid particles at different time instants.
The initially circular droplet gradually deforms under the influence of shear flow and eventually reaches a stable elliptical shape due to the interplay between viscous forces and surface tension.
Fig. \ref{figs:2d_drop_shear_D} illustrates the relationship between the deformation parameter $D$ and the capillary number $Ca$ obtained from the numerical simulation in this study. 
For comparison, the analytical results by Taylor \cite{taylor1934formation} and the SPH results from Adami et al. \cite{adami2010new} are also included.
The results demonstrate good agreement between the theoretical and previous numerical values, with a maximum error of around 5.4\%.

\begin{figure}[htb!]
	\centering
	\includegraphics[trim = 0cm 0cm 0cm 0cm, clip,width=1\textwidth]{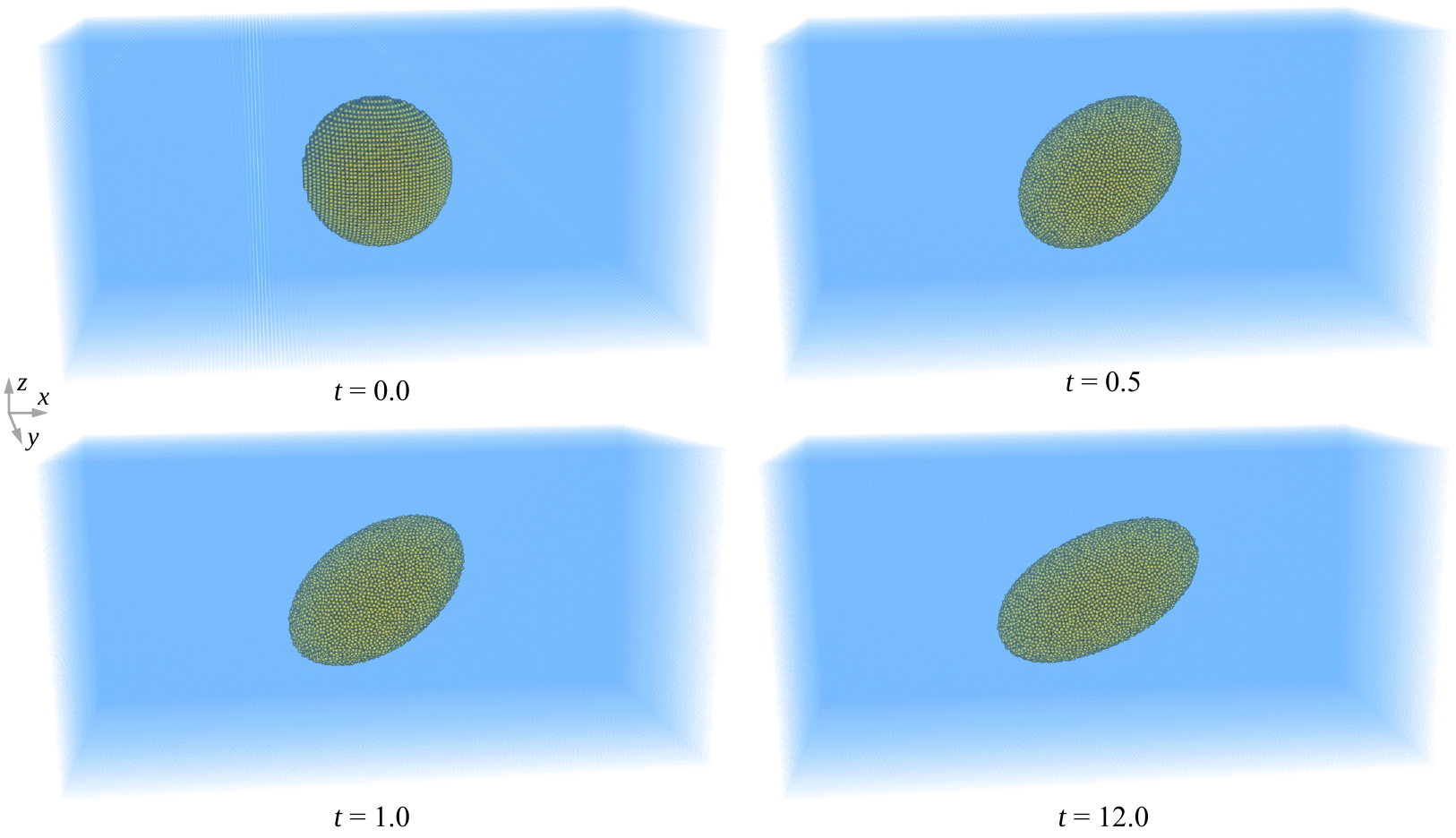}
	\caption{3D drop deformation in shear flow: positions of droplet particles and shearing fluid particles at $t=0.0$, 0.5, 1.0 and 12.0. Here, $Ca=0.25$, $Re=1.0$, and $R/dp=20$.}
	\label{figs:3d_drop_shear_snapshot}
\end{figure}

We further extend the aforementioned case to three dimensions to demonstrate the stability of the proposed algorithm in 3D multiphase flows. 
A spherical droplet with a radius of $R=1$ is placed at the center of the computational domain, which has dimensions of $8R\times 4R\times 4R$. The velocities of the top and bottom walls are set to $\pm 2$, respectively, while periodic boundary conditions are applied to the remaining walls. The density ratio and viscosity ratio of the two fluids are both set to 1, and the particle spacing is $dp=R/20=0.05$. 
The Reynolds number is 1, and the capillary number is 0.25. 
Fig.~\ref{figs:3d_drop_shear_snapshot} illustrates the state of the droplet at different time instances during the simulation. Similar to the 2D scenario, the droplet undergoes elongation and subsequently reaches an equilibrium state under the combined effects of surface tension and viscous forces. 
Throughout the simulation, the computation remains stable, and the particle distribution remains uniform, demonstrating the applicability of the proposed method to realistic 3D problems.
%%%%%%%%%%%%%%%%%%%%%%%%%%%%%%%%%%%%%%%%%%%%%%%%%%%%%%%%%%%%%
% 5.3 Square droplet oscillation
%%%%%%%%%%%%%%%%%%%%%%%%%%%%%%%%%%%%%%%%%%%%%%%%%%%%%%%%%%%%%
\subsection{Square droplet oscillation}
\label{square-droplet-oscillation}
\begin{figure}[htb!]
	\centering
	\includegraphics[trim = 0cm 0cm 0cm 0cm, clip,width=0.5\textwidth]{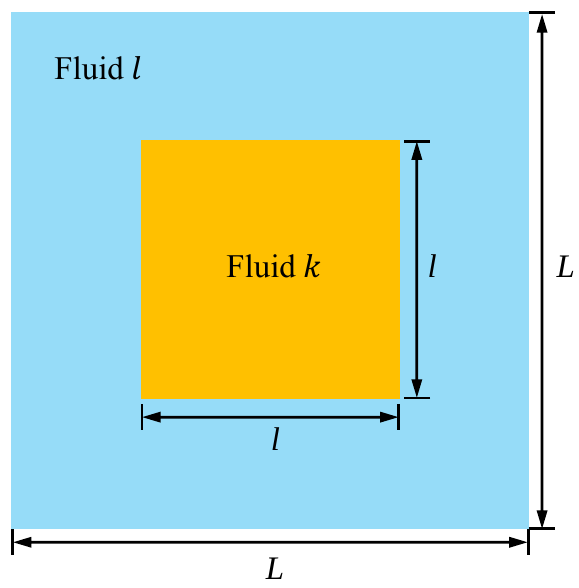}
	\caption{2D square droplet oscillation: model setup.}
	\label{figs:2d_square_drop_setup}
\end{figure}

To further validate the accuracy of the surface tension formulation presented in this study, a test case involving the motion of a square droplet under the influence of surface tension is examined \cite{adami2010new,zhang2023improved,yang2022consistent,blank2023modeling}. As shown in Fig.~\ref{figs:2d_square_drop_setup}, an initially square-shaped droplet is placed in a zero-gravity environment surrounded by another fluid, with dimensions $L=2.0$ and $l=1.0$. The density ratio is set to $\rho_k / \rho_l=1000$, the viscosity ratio to $\mu_k / \mu_l=100$, and the fluid properties are defined as $\rho_k =1.0$ and $\mu_k =0.05$. The surface tension coefficient is assigned a value of $\gamma =1.0$.

\begin{figure}[htb!]
	\centering
	\includegraphics[trim = 0cm 0cm 0cm 0cm, clip,width=1\textwidth]{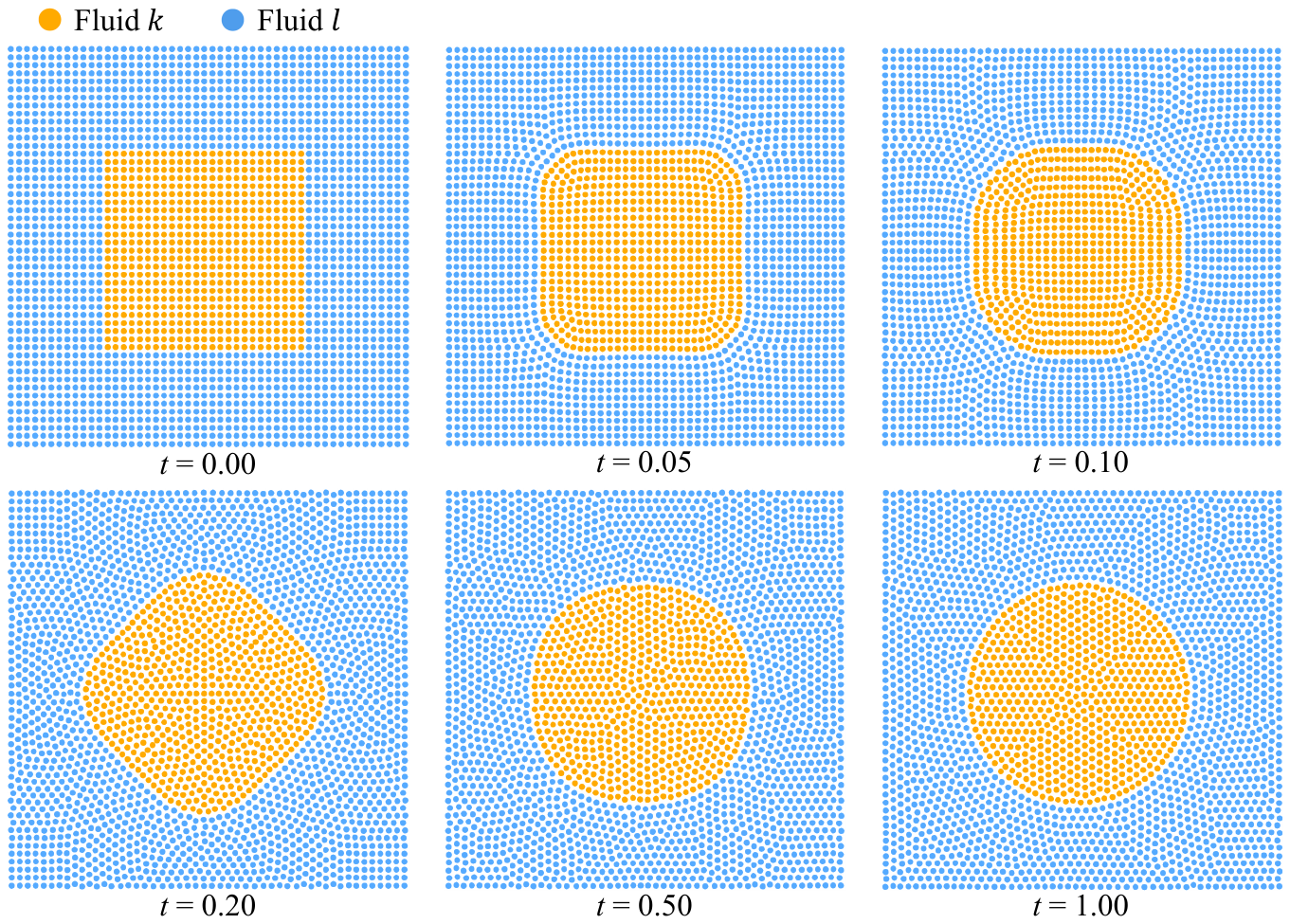}
	\caption{2D square droplet oscillation: time evolution of the 2D square droplet at $t=0.00$, 0.05, 0.10, 0.20, 0.50 and 1.00. Here, $\rho_k / \rho_l=1000$ and $dp=0.02$.}
	\label{figs:2d_square_drop_snapshot}
\end{figure}

Fig. \ref{figs:2d_square_drop_snapshot} illustrates the configurations of the droplet at various simulated time points. Driven by surface tension, the fluid-fluid interface undergoes continuous deformation until the interfacial energy is minimized, ultimately achieving a circular shape at equilibrium.

\begin{figure}[htb!]
	\centering
	\includegraphics[trim = 0cm 0cm 0cm 0cm, clip,width=0.5\textwidth]{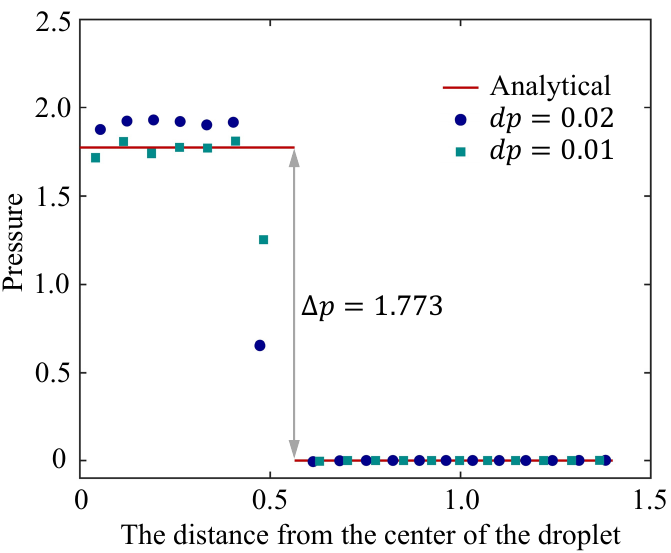}
	\caption{2D square droplet oscillation: pressure as a function of distance from the droplet center at steady state. $\rho_k / \rho_l=1000$.}
	\label{figs:2d_square_drop_pressure}
\end{figure}

At steady state, the pressure difference between the inner and outer fluids can be characterized by the Young-Laplace equation.
In two dimensions, the pressure jump across the interface must satisfy the following condition
\begin{equation}
   \Delta p =  \frac{\gamma}{R} = \frac{\gamma\sqrt{\pi } }{l}
  \label{young-laplace}
\end{equation}
where $R$ is the radius of the circular droplet at equilibrium.
Fig.~\ref{figs:2d_square_drop_pressure} illustrates the variation of pressure as a function of distance from the droplet center at steady state under different resolutions, with the analytical solution included for comparison. 
As can be observed, the trend of pressure variation aligns well with the analytical result. 
Moreover, at the higher resolution $(dp=0.01)$, the pressure values exhibit close agreement with the theoretical predictions, further validating the accuracy of the present formulation.

\begin{figure}[htb!]
	\centering
	\includegraphics[trim = 0cm 0cm 0cm 0cm, clip,width=1\textwidth]{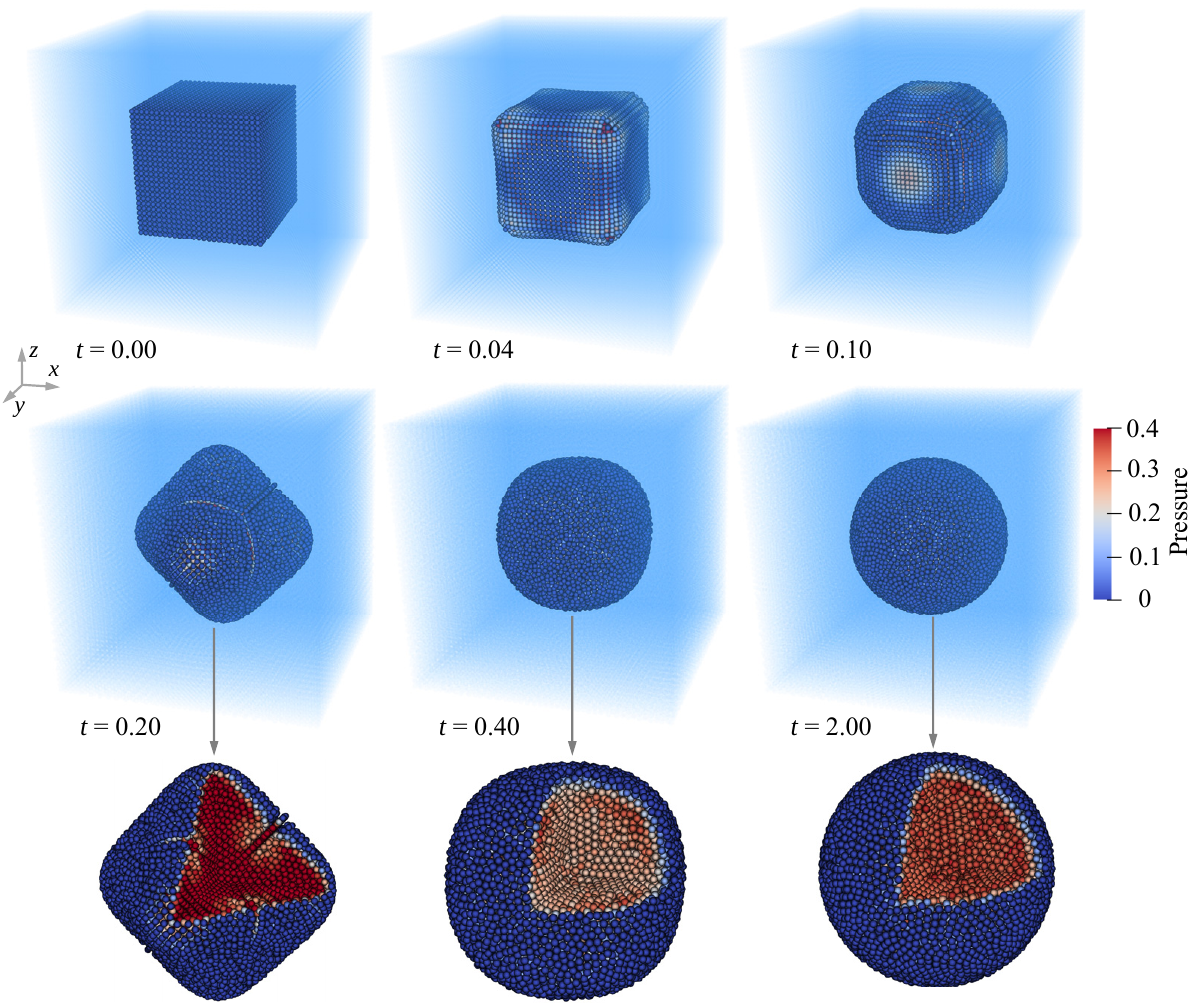}
	\caption{3D cubic droplet oscillation: time evolution of the 3D square droplet at $t=0.00$, 0.04, 0.10, 0.20, 0.40 and 2.00. Here, $\rho_k / \rho_l=1000$ and $dp=0.02$.}
	\label{figs:3d_square_drop_snapshot}
\end{figure}

Next, the case is extended to a 3D scenario. A cubic liquid droplet with a side length of 1 is placed within another fluid, with the physical parameters of the fluids remaining consistent with those in the 2D case. 
Fig.~\ref{figs:3d_square_drop_snapshot} depicts the state of the droplet at various simulated time points. 
Similar to the 2D case, the sharp corners of the cube initially experience significant surface tension, causing them to be compressed inward. 
After undergoing a series of dynamic deformations, the droplet eventually reaches equilibrium. 
At equilibrium, the particles at the droplet interface are distributed uniformly, demonstrating the effectiveness of the surface tension algorithm proposed in this study. 
Fig.~\ref{figs:3d_square_drop_snapshot} also illustrates the pressure distribution throughout the process. The bottom row of Fig.~\ref{figs:3d_square_drop_snapshot} presents a cross-sectional view of the droplet, revealing the internal pressure distribution.
%%%%%%%%%%%%%%%%%%%%%%%%%%%%%%%%%%%%%%%%%%%%%%%%%%%%%%%%%%%%%
% 5.4 High-velocity droplet impact
%%%%%%%%%%%%%%%%%%%%%%%%%%%%%%%%%%%%%%%%%%%%%%%%%%%%%%%%%%%%%
\subsection{High-velocity droplet impact}
\label{droplet-impact}
\begin{figure}[htb!]
	\centering
	\includegraphics[trim = 0cm 0cm 0cm 0cm, clip,width=0.5\textwidth]{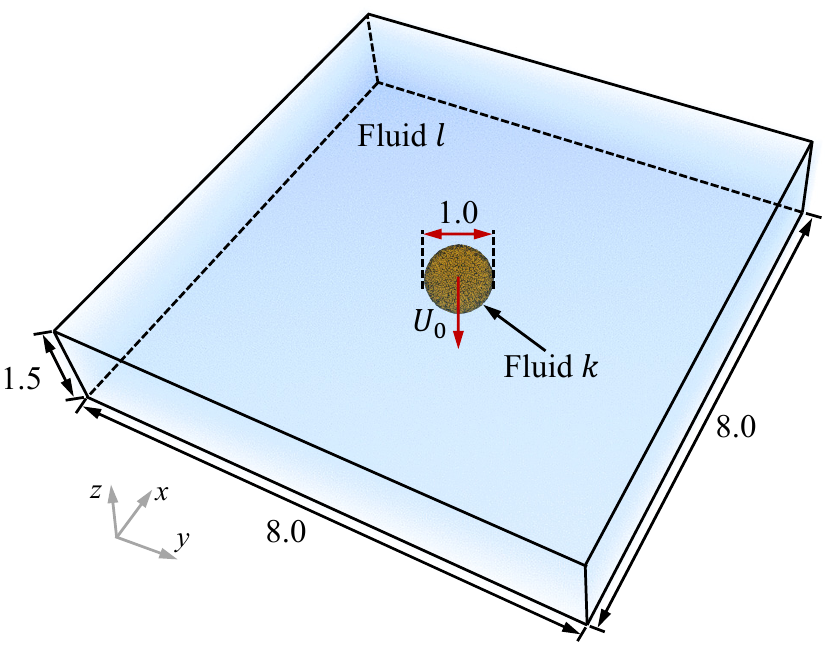}
	\caption{High-velocity droplet impact: model setup.}
	\label{figs:drop_impact_setup}
\end{figure}

This section further demonstrates the effectiveness of the proposed surface tension formulation under conditions of high Reynolds and Weber numbers.
The model setup is shown in Fig. \ref{figs:drop_impact_setup}: a spherical droplet (fluid $k$) with a diameter of $d = 1$ is encapsulated in another fluid $l$, which has significantly lower density and viscosity (such as air).
The density ratio between fluid $k$ and fluid $l$ is $\rho_k / \rho_l=1000$, and the viscosity ratio is $\mu_k / \mu_l=100$, with $\rho_k=1$ and $\mu_k=0.005$. The surface tension coefficient is $\gamma=0.1$.
The computational domain is of size $8 \times 8 \times 1.5$. 
Initially, the droplet is placed at the center along the $x$ and $y$ axes, and its bottom is in direct contact with the wall boundary along the $z$ axis.
The droplet moves toward the negative z-axis with an initial velocity of $U_0$ and impacts the wall. The Reynolds number and Weber number are defined as
\begin{equation}
   Re = \frac{\rho_k U_0 d}{\mu_k}, \quad We = \frac{\rho_k U_0^2 d}{\gamma}
  \label{Re-We}
\end{equation}
By varying only the initial velocity ($U_0=0.5, 1, 5, 10, 50$) of the droplet while keeping all other parameters unchanged, results under different Reynolds numbers and Weber numbers are obtained.

\begin{figure}[htb!]
	\centering
	\includegraphics[trim = 0cm 0cm 0cm 0cm, clip,width=1\textwidth]{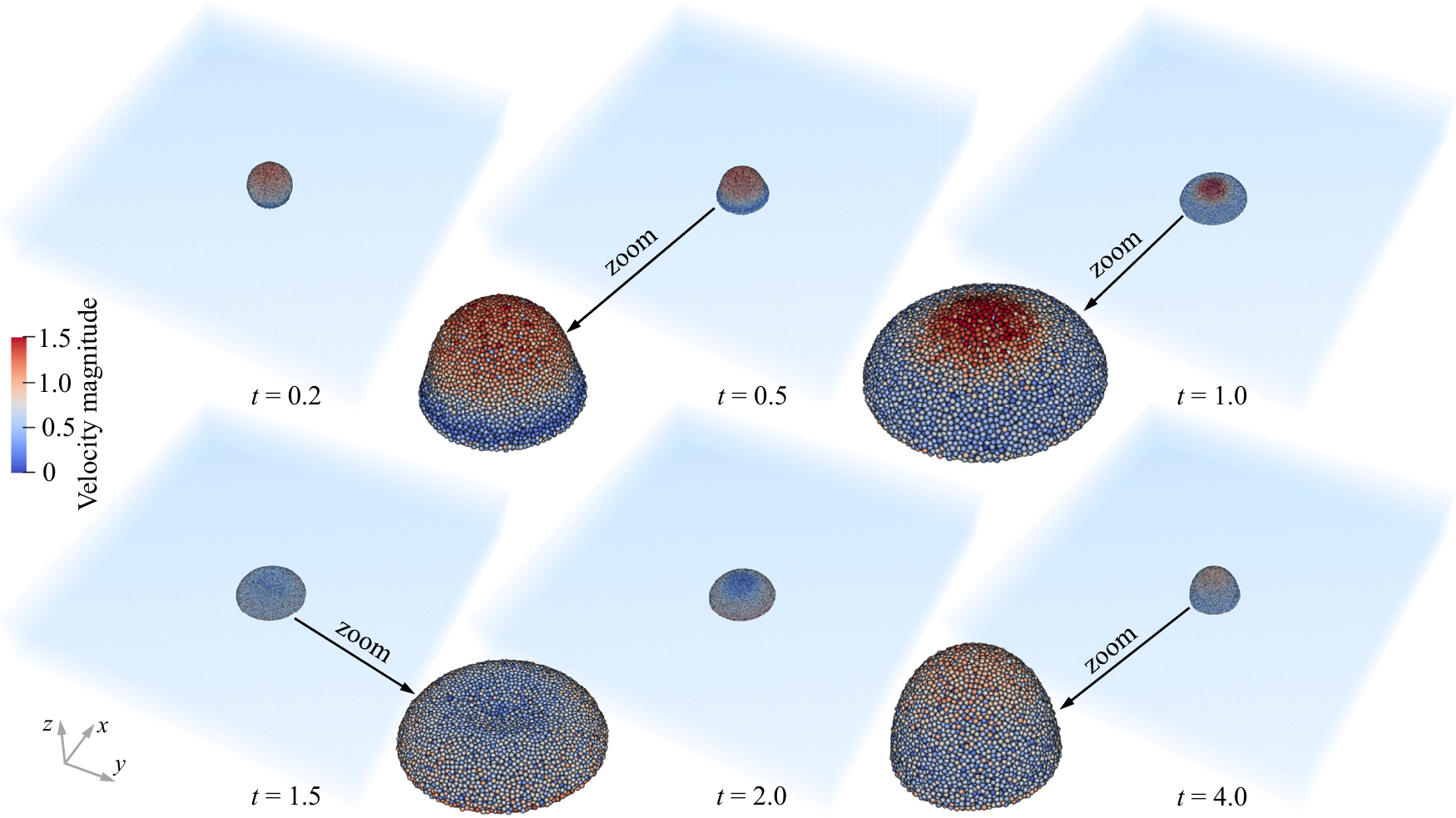}
	\caption{High-velocity droplet impact: time evolution of the droplet at $t=0.2$, 0.5, 1.0, 1.5, 2.0 and 4.0. Here, $U_0=0.5$, $Re=100$, and $We=2.5$.}
	\label{figs:drop_impact_Re_100}
\end{figure}

\begin{figure}[htb!]
	\centering
	\includegraphics[trim = 0cm 0cm 0cm 0cm, clip,width=1\textwidth]{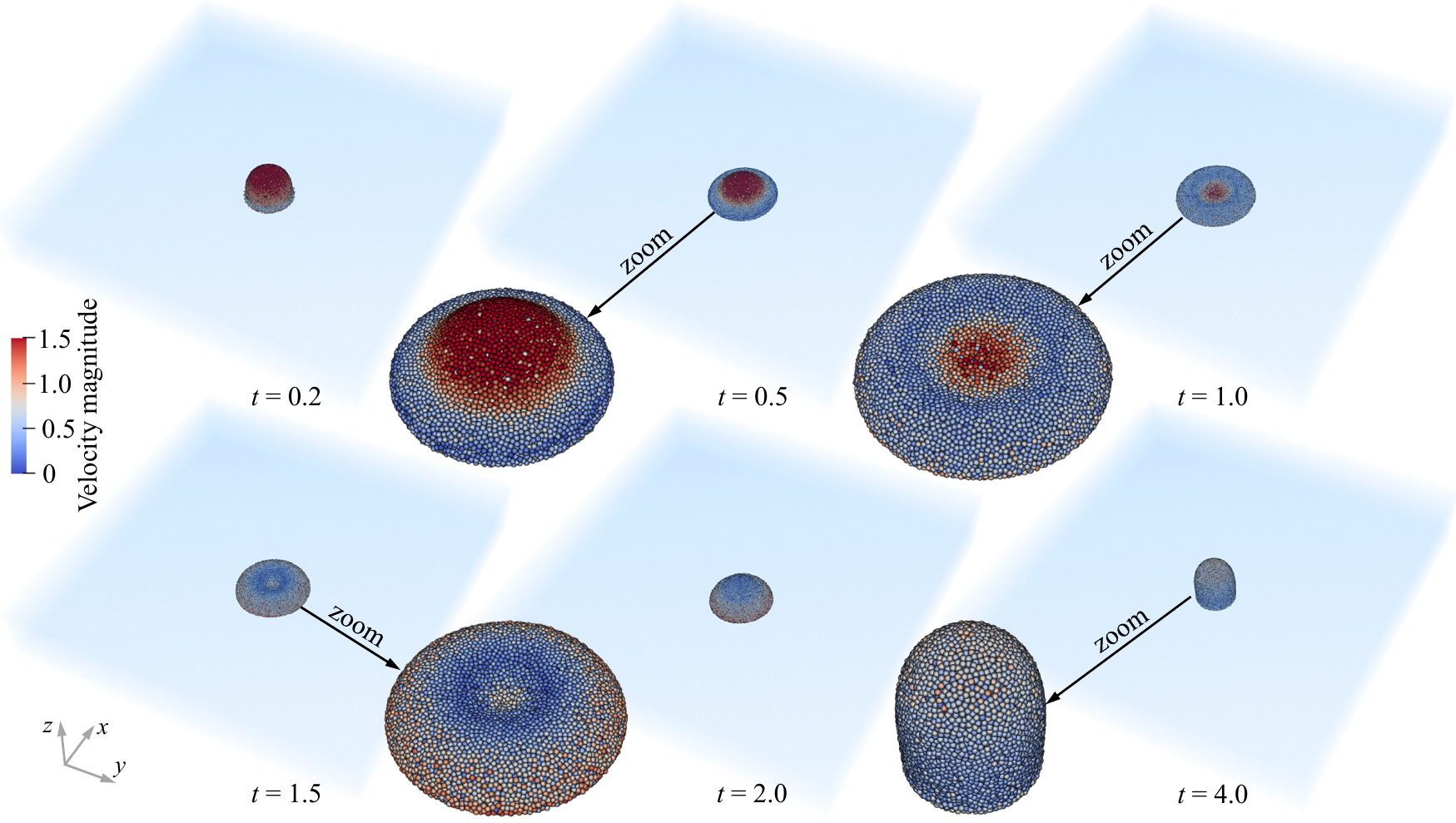}
	\caption{High-velocity droplet impact: time evolution of the droplet at $t=0.2$, 0.5, 1.0, 1.5, 2.0 and 4.0. Here, $U_0=1$, $Re=200$, and $We=10$.}
	\label{figs:drop_impact_Re_200}
\end{figure}

The results at lower Reynolds numbers and Weber numbers are presented in Fig. \ref{figs:drop_impact_Re_100} and Fig. \ref{figs:drop_impact_Re_200}, respectively.
At $Re = 100$ and $We = 2.5$ (Fig. \ref{figs:drop_impact_Re_100}), the droplet undergoes gentle spreading upon impact ($t=0.5$). The simulation reveals a smooth radial expansion, with the droplet forming a pancake-like shape ($t=1.5$) at its maximum spreading radius. No evidence of fingering or splashing is observed, which is consistent with the experimental observations \cite{visser2012microdroplet} for low-Weber-number impacts. 
The low inertial forces relative to surface tension ensure that the spreading process is dominated by capillary forces. Subsequently, under the influence of surface tension, the droplet gradually retracts and exhibits a rebound phenomenon.

When the $Re$ increases to 200 and the $We$ increases to 10 (Fig. \ref{figs:drop_impact_Re_200}), the results exhibit more pronounced spreading dynamics.
The droplet expands radially and deforms into a disc-like structure ($t=1.0$), forming a thin liquid sheet with a thicker rim \cite{visser2012microdroplet}. 
Fingering instabilities remain absent due to the relatively low $We$, and the droplet maintains a smooth profile during spreading. 
The progression is still governed by capillary effects, as seen in the transition regime described by Clanet et al. \cite{clanet2004maximal}.
Due to the greater initial kinetic energy, the rebound height of the droplet at the final time ($t=4.0$) is also higher compared to Fig. \ref{figs:drop_impact_Re_100}.

\begin{figure}[htb!]
	\centering
	\includegraphics[trim = 0cm 0cm 0cm 0cm, clip,width=1\textwidth]{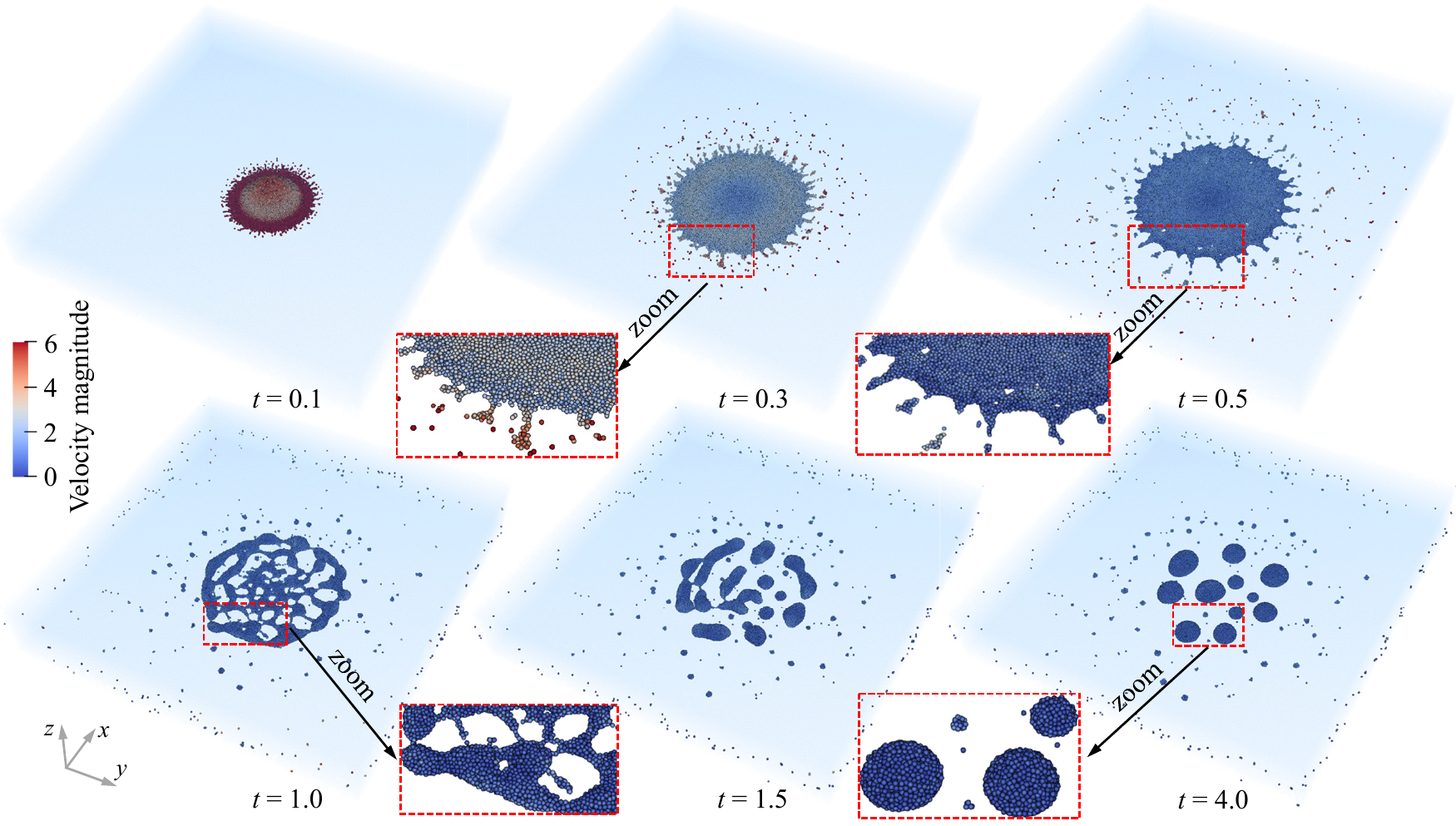}
	\caption{High-velocity droplet impact: time evolution of the droplet at $t=0.1$, 0.3, 0.5, 1.0, 1.5 and 4.0. Here, $U_0=5$, $Re=1000$, and $We=250$.}
	\label{figs:drop_impact_Re_1000}
\end{figure}

At high Reynolds and Weber numbers ($Re=1000$, $We=250$), as shown in Fig. \ref{figs:drop_impact_Re_1000}, the droplet impact dynamics become more complex.
The droplet exhibits distinct fingering instabilities at the rim of the spreading lamella ($t=0.3$). 
These fingers grow and elongate, eventually breaking off to form satellite droplets ($t=0.5$), consistent with experimental findings \cite{mehdizadeh2004formation, pan2010breakup}. 
The Weber number is sufficiently large to overcome surface tension, leading to splashing behavior. 
When the droplet spreading reaches a certain extent, the non-splashing main body becomes exceedingly thin. 
At this stage, as the kinetic energy gradually dissipates, the lamella in certain regions ruptures ($t=1.0$) due to the combined effects of surface tension and the retraction of the lamella, leading to the formation of a net-like structure. Subsequently, under the influence of surface tension, the adjacent liquid gradually coalesces ($t=1.5$), forming smaller droplets ($t=4.0$).

\begin{figure}[htb!]
	\centering
	\includegraphics[trim = 0cm 0cm 0cm 0cm, clip,width=1\textwidth]{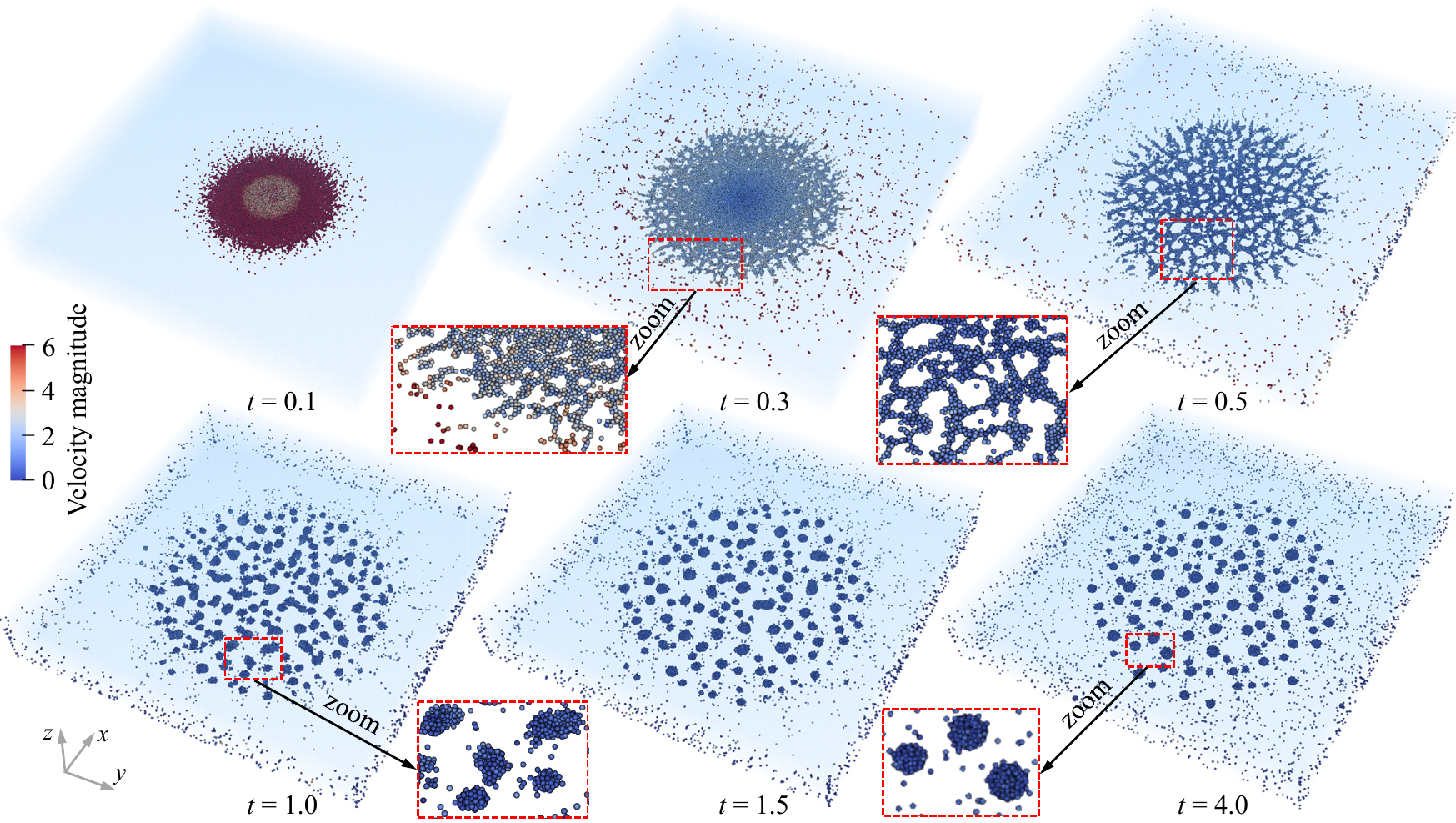}
	\caption{High-velocity droplet impact: time evolution of the droplet at $t=0.1$, 0.3, 0.5, 1.0, 1.5 and 4.0. Here, $U_0=10$, $Re=2000$, and $We=1000$.}
	\label{figs:drop_impact_Re_2000}
\end{figure}

With a further increase in $Re$ and $We$ ($Re=2000$, $We=1000$), as shown in Fig. \ref{figs:drop_impact_Re_2000}, the droplet impact results in extensive splashing, characterized by the detachment of numerous satellite droplets from the rim ($t=0.1$). 
The simulation demonstrates the formation of irregular liquid fragments, indicative of a fully splashing regime ($t=0.5$). The dynamics observed are in agreement with the high-Weber-number experiments \cite{pan2010breakup}.
The lamella becomes increasingly unstable, and due to the larger initial kinetic energy, the droplet spreads more extensively compared to Fig. \ref{figs:drop_impact_Re_1000}. 
As a result, it breaks into a greater number of smaller droplets, which eventually coalesce and reach a stable state ($t=4.0$).

\begin{figure}[htb!]
	\centering
	\includegraphics[trim = 0cm 0cm 0cm 0cm, clip,width=1\textwidth]{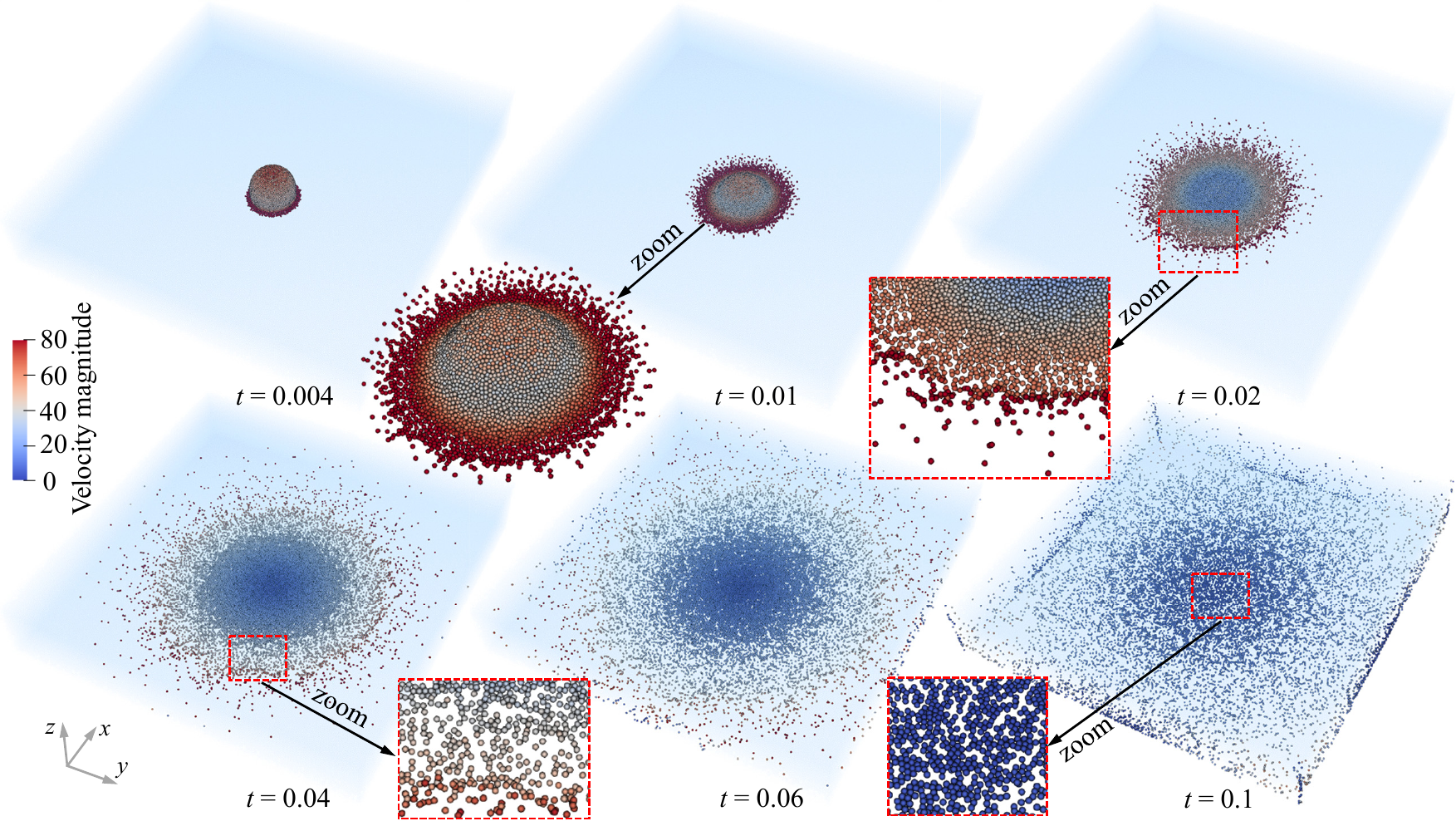}
	\caption{High-velocity droplet impact: time evolution of the droplet at $t=0.004$, 0.01, 0.02, 0.04, 0.06 and 0.1. Here, $U_0=50$, $Re=10000$, and $We=25000$.}
	\label{figs:drop_impact_Re_10000}
\end{figure}

At extremely high Reynolds and Weber numbers ($Re=10000$, $We=25000$), as shown in Fig. \ref{figs:drop_impact_Re_10000}, the lamella rapidly disintegrates into a complex spray of fragments and satellite droplets ($t=0.02$).
Due to the extremely high initial velocity, the droplet spreads rapidly outward along the wall and does not stop even as it approaches the boundaries of the computational domain. 
At the final time ($t=0.1$), the liquid forms a thin layer closely adhering to the wall.
Even under such high Reynolds and Weber numbers, no numerical instability is observed throughout the entire computation process.
%%%%%%%%%%%%%%%%%%%%%%%%%%%%%%%%%%%%%%%%%%%%%%%%%%%%%%%%%%%%%
%
% 6 Conclusion
%
%%%%%%%%%%%%%%%%%%%%%%%%%%%%%%%%%%%%%%%%%%%%%%%%%%%%%%%%%%%%%
\section{Conclusions}
\label{conclusions}
This study proposes a computational framework based on the Riemann-SPH method for simulating surface tension in multiphase flows with high density and viscosity ratios. 
The surface tension is calculated and discretized based on surface stress to ensure momentum conservation. 

This work is the first to thoroughly investigate the underlying causes of particle disorder frequently observed at interfaces between different fluids. 
It is identified that such disorder originates from a type of numerical instability, defined in this work as zero-surface-energy modes, which occur at fluid-fluid interfaces.
Specifically, during the computation of the color gradient, the contributions of symmetrically positioned neighboring particles cancel each other out, leading to an underestimation of the surface tension force. 
To address this issue, a penalty force is introduced to compensate for the inaccurately underestimated surface tension. 
Importantly, this correction term is also momentum-conserving and requires no parameter tuning.

The stability and convergence of the proposed surface tension calculation method are validated through several benchmark cases with known theoretical solutions. 
Finally, a 3D high-velocity droplet impact simulation is performed to demonstrate the robustness of the proposed computational framework under high Reynolds and Weber numbers ($Re=10000$, $We=25000$), marking the first successful application of particle-based methods to multiphase flow simulations under such extreme conditions.
A qualitative comparison with experimental results further confirms the reliability of the numerical predictions.
%%%%%%%%%%%%%%%%%%%%%%%%%%%%%%%%%%%%%%%%%%%%%%%%%%%%%%%%%%%%%
%
% Section
%
%%%%%%%%%%%%%%%%%%%%%%%%%%%%%%%%%%%%%%%%%%%%%%%%%%%%%%%%%%%%%
\section*{CRediT authorship contribution statement}
\addcontentsline{toc}{section}{CRediT}

Shuaihao Zhang: Conceptualization, Methodology, Investigation, Visualization, Validation, Formal analysis, Writing - original draft, Writing - review \& editing. 
Sérgio D.N. Lourenço: Supervision, Writing - review \& editing, Funding acquisition.
Xiangyu Hu: Supervision, Methodology, Investigation, Writing - review \& editing.
%%%%%%%%%%%%%%%%%%%%%%%%%%%%%%%%%%%%%%%%%%%%%%%%%%%%%%%%%%%%%
%
% Section
%
%%%%%%%%%%%%%%%%%%%%%%%%%%%%%%%%%%%%%%%%%%%%%%%%%%%%%%%%%%%%%
\section*{Declaration of competing interest}
\addcontentsline{toc}{section}{declaration-interest}

The authors declare that they have no known competing financial interests or personal relationships that could
have appeared to influence the work reported in this paper.
%%%%%%%%%%%%%%%%%%%%%%%%%%%%%%%%%%%%%%%%%%%%%%%%%%%%%%%%%%%%%
%
% Section
%
%%%%%%%%%%%%%%%%%%%%%%%%%%%%%%%%%%%%%%%%%%%%%%%%%%%%%%%%%%%%%
\section*{Acknowledgements}
\addcontentsline{toc}{section}{acknowledgement}

Xiangyu Hu would like to express his gratitude to the German Research Foundation (DFG) for their sponsorship of this research under grant number DFG HU1527/12-4.
Sérgio D.N. Lourenço would like to express his gratitude to the Research Grants Council Hong Kong for their sponsorship of this research under a Collaborative Research Fund (C6006-20GF).
The computations were performed using research computing facilities offered by Information Technology Services, the University of Hong Kong.
%%%%%%%%%%%%%%%%%%%%%%%%%%%%%%%%%%%%%%%%%%%%%%%%%%%%%%%%%%%%%
%
% Section
%
%%%%%%%%%%%%%%%%%%%%%%%%%%%%%%%%%%%%%%%%%%%%%%%%%%%%%%%%%%%%%
\bibliographystyle{elsarticle-num}
\bibliography{surface_tension}
%%%%%%%%%%%%%%%%%%%%%%%%%%%%%%%%%%%%%%%%%%%%%%%%%%%%%%%%%%%%%
%
% Section
%
%%%%%%%%%%%%%%%%%%%%%%%%%%%%%%%%%%%%%%%%%%%%%%%%%%%%%%%%%%%%%
\end{document}